\documentclass[pra, aps,
twocolumn,
amsmath,amssymb
,floatfix
]{revtex4-1}

\usepackage{physics}
\usepackage{epsf}
\usepackage{float}
\usepackage{graphicx}
\usepackage{amsmath}
\usepackage[usenames]{color}
\usepackage{float}
\usepackage{bm}
\usepackage{relsize}
\usepackage[normalem]{ulem}

\def \be {\begin{equation}}
\def \ee {\end{equation}}
\def \bea {\begin{align}}
\def \eea {\end{align}}
\def \p {\partial}
\def \BEA {\begin{eqnarray}}
\def \EEA {\end{eqnarray}}
\def \BC {\begin{cases}}
	\def \EC {\end{cases}}

\usepackage{siunitx}
\usepackage{tabularx}
\usepackage{xcolor}
\usepackage[version=4]{mhchem}
\usepackage{ctable}

\usepackage[english]{babel}

\usepackage[unicode=true,colorlinks=true,citecolor=blue,urlcolor=blue]{hyperref}

\def \be {\begin{equation}}
\def \ee {\end{equation}}
\def \bea {\begin{align}}
\def \eea {\end{align}}
\def \p {\partial}
\def\bee{\begin{eqnarray}}
\def\eee{\end{eqnarray}}
\def \BC {\begin{cases}}
	\def \EC {\end{cases}}
\def \m {\bm }

\newcommand{\EMs}[1]{\textcolor{cyan}{\sout{#1}}}

\newcommand{\re}[1]{\textcolor{red}{#1}}

\newcommand{\Degree}[1]{\SI{#1}{\degree}}

\newenvironment{comment}{}{}

\begin{document}

\title{Cyclotron- and magnetoplasmon resonances in bilayer graphene ratchets}

\author{E. M{\"o}nch$^1$, S.~O. Potashin$^2$,  K.~Lindner$^1$, I. Yahniuk$^{1,3}$, L.~E. Golub$^{1,2}$, V.~Yu. Kachorovskii$^{2,3}$, V.~V.~Bel'kov$^{2}$, 	
	R.~Huber$^1$, K.~Watanabe$^4$, T.~Taniguchi$^5$, J.~Eroms$^1$, D.~Weiss$^1$,  and S.~D.~Ganichev$^{1,3}$}

\affiliation{$^1$Terahertz Center, University of Regensburg, 93040 Regensburg, Germany}

\affiliation{$^2$Ioffe Institute, 194021 St. Petersburg, Russia}

\affiliation{$^3$CENTERA, Institute of High Pressure Physics PAS, 01142 Warsaw, Poland}

\affiliation{$^4$Research Center for Functional Materials, 	National Institute for Materials Science, 1-1 Namiki, Tsukuba 305-0044, Japan}

\affiliation{$^5$International Center for Materials Nanoarchitectonics, National Institute for Materials Science,  1-1 Namiki, Tsukuba 305-0044, Japan}


\begin{abstract}
	 We report on a tunable --- by magnetic field and gate voltage --- conversion  of terahertz radiation into a dc current in spatially modulated bilayer graphene. We experimentally demonstrate that the underlying physics is related to the so-called ratchet effect. Our key findings are the direct observation of a  sharp cyclotron resonance in the photocurrent and the demonstration of two effects caused by electron-electron interaction: the plasmonic splitting of the resonance due to long-range Coulomb coupling and the partial suppression of its second harmonic due to fast interparticle collisions. We  develop a theory which perfectly fits our data. We  argue  that the ratchet current is generated in the    
 hydrodynamic  regime of non-ideal electron liquid.
\end{abstract}

\maketitle

\section{Introduction}
\label{intro}

One of the most general and fascinating phenomena in optoelectronics is the ratchet effect --- the generation of a dc electric current in response to an ac electric field in systems with broken inversion symmetry~\cite{Haenggi2009,Ivchenko2011,Denisov2014, Bercioux2015,Reichhardt2017,Ganichev2017}.
This general definition applies both for short-channel devices, like field effect transistors with asymmetric boundary conditions~\cite{Vicarelli2012,Muraviev2013,Cai2015,Wang2015, Bandurin2018a,Rupper2018,Tomadin2013a,Spirito2014, Auton2017}
and for long periodic grating gate structures with an asymmetric configuration of the gate electrodes~\cite{Olbrich2009ratchet,Olbrich2011,Nalitov2012,Bisotto2011, Otsuji2013,Kurita2014,BoubangaTombet2014, Faltermeier2015,Olbrich2016,Popov2016, Ivchenko2011,Fateev2019,BoubangaTombet2020, DelgadoNotario2020}.
While the ratchet effect has been studied in various low-dimensional systems based on GaAs~\cite{Olbrich2009ratchet,Olbrich2011,Ivchenko2011}, SiGe~\cite{Bisotto2011} and InGaAs heterostructures~\cite{Kurita2014,BoubangaTombet2014, Faltermeier2015} and graphene~\cite{Nalitov2012,Otsuji2013,Olbrich2016, Fateev2019,BoubangaTombet2020,DelgadoNotario2020}, in particular in magnetic fields~\cite{Kannan2011,Kannan2012,Drexler2013, Faltermeier2017, Faltermeier2018,Hubmann2020,Sai2021},
the cyclotron resonance (CR) has  not been observed so far. Therefore, the interplay of CR and plasmonic effects could not yet be investigated.

Here we report on the observation of two resonant ratchet effects caused by  CR and magnetoplasmon (MP) resonances. Both effects are observed by studying the  conversion of terahertz (THz) radiation into a dc current in bilayer graphene (BLG) superimposed with a lateral superlattice  consisting of a dual-grating top gate (DGG) structure. The resonances are observed in the Shubnikov-de Haas (SdH) regime, where the ratchet current exhibits sign-alternating magneto-oscillations, with greatly enhanced amplitudes as compared to the photosignal of the  zero magnetic field ratchet effect previously studied in BLG DGG in Ref.~\cite{Moench2022}. We develop a theory which fully describes all experimental findings. Remarkably, CR and MP resonances have different magnetic field positions. This is a characteristic feature of the ratchet effect being in a sharp contrast to results for conventional two- dimensional structures where plasmonic effects do not lead to a splitting of CR but only to a shift of the CR position. Physically, plasmonic splitting of the CR is due to the spatial modulation of the incoming radiation and to the non-linear nature of the ratchet effect. Most importantly, the radiation field has a homogeneous component, causing the CR and a component modulated with finite wavevector $q=2\pi/L$ ($L$ is the superlattice period), which leads to the MP resonance. Non-linear mixing of these two components results in the interference contribution which contains two resonances.

Furthermore, we observe an enhancement of the SdH-related oscillations in the ratchet current at magnetic fields corresponding to one half of the CR position ($\omega=2\omega_c)$. Our theory reveals that the amplitude of this second harmonic is determined by the relaxation rate of the second angular harmonic in  the velocity distribution function. This rate is strongly enhanced in the hydrodynamic (HD) regime, where  electron-electron (ee) collisions dominate over impurity scattering. To explore the interplay between HD and impurity-dominated or so-called drift-diffusion (DD) regimes we develop the theory for two  cases: $\gamma \ll \gamma_{\rm ee}$ and  $\gamma \gg \gamma_{\rm ee}$, where $\gamma_{\rm ee}$ and $\gamma$ are ee- and electron-impurity scattering rates, respectively. We argue that our results at liquid He temperature suggest that our system is in the HD regime. This is in excellent agreement with our recent study of the same structures at zero magnetic field \cite{Moench2022}. The key justification that the system is very close to the HD regime is the fairly small amplitude of the second harmonic as compared to the first one. Nevertheless, although being small, the second harmonic is clearly seen in the experiment implying that we are dealing with a non-ideal electron fluid with finite viscosity.

The paper is organized as follows. In Sec.~\ref{state of the art} we introduce the basic notions of the ratchet effect and discuss the state of the art regarding the physics of ratchet currents. In Sec.~\ref{samples-methods} we describe the investigated samples and the experimental technique. In Sec.~\ref{results} we discuss the experimental results. In the following Secs.~\ref{model} and \ref{theory_results} we present the model approach and theoretical results, respectively. In Sec.~\ref{discussion} we discuss the results and compare experimental and theoretical magnetic field dependencies of the ratchet effect. Finally, in Sec.~\ref{summary} we summarize the results.

\section{Ratchet effect: Basic notions and state of the art}
\label{state of the art}
We start with recalling some basic notions of the ratchet effect and discussing the state of the art. In the periodic grating gate structures, which we study in the current work, both the density of electrons and the field amplitude of the incoming radiation 
$\bm E(x,t) = \bm E_0(x) \text{e}^{-i\omega t} + \bm E_0^*(x) \text{e}^{i\omega t}$ are modulated. 
The former one is due to the electrostatic potential $V(x),$ which can be manipulated  by   two top gates  of the DGG,   and the latter one is  a result of near-field effects   caused by the THz radiation propagating through the grating. A nonzero dc current appears if the structure possesses an asymmetry, so that the asymmetry parameter
\begin{equation}
\label{Xi0}
\Xi=\overline{ {dV\over dx}|\bm E_0(x)|^2}
\end{equation}
is nonzero~\cite{Ivchenko2011}. 
Here averaging is taken over the $x$-coordinate along the spatial modulation.
Although weak built-in asymmetry can be  present, in principle,  in an unbiased structure, it can be drastically enhanced changing also its sign by biasing the top gates in an asymmetric configuration. The ratchet effect was treated theoretically and observed experimentally in various bulk crystals and low-dimensional structures both at zero~\cite{Olbrich2011,Nalitov2012,Otsuji2013, Kurita2014,BoubangaTombet2014,Faltermeier2015, Olbrich2016,Popov2016,Ganichev2017,Fateev2019, BoubangaTombet2020,DelgadoNotario2020}
and non-zero~\cite{Faltermeier2017,Faltermeier2018,Hubmann2020, Sai2021}
magnetic field. However, some basic issues of this effect still remain puzzling. One of the major unsolved issues is the role of the ee-interaction. The standard calculations of the ratchet effect~\cite{Ivchenko2011} were performed in DD approximation where the ee-interaction was ignored. Such a ratchet effect is sometimes referred to as electronic ratchet. Actually, the effect of the ee-interaction is  twofold and can be quite strong. First of all, sufficiently fast ee-collisions can drive the system into the HD regime. Secondly, ee-interaction controls the plasmonic oscillations, so that a new frequency scale, the plasma frequency, $\omega_q$ appears in the problem, where $q$ is the inverse characteristic length scale of the system. For a device with short length, for example, a single field-effect transistor (FET), $q$ is proportional to the inverse length of the device. For periodic-gate structures it is $q= 2\pi/L$. The plasmonic ratchet effect is dramatically enhanced in the vicinity of the plasmonic resonances both for a single FET with asymmetric boundary conditions~\cite{Dyakonov1996,Echtermeyer2011,Vicarelli2012,Muraviev2013,Cai2015,Wang2015,Bandurin2018a,Rupper2018} and for the periodic asymmetric DGG~\cite{Popov2011,Watanabe2013b, Rozhansky2015,Spisser2015,Olbrich2016,Fateev2017, Fateev2017a,Yu2018,Fateev2019,DelgadoNotario2020,Bai2021,Bai2022}.
Historically, electronic ratchet effects are treated theoretically within the  DD approximation, while  HD is usually used to describe plasmonic effects. The plasmonic ratchet can be turned off by using high excitation frequencies, $\omega \gg \omega_q.$  However, even in this case, one needs to choose between DD or HD approximation depending on the relation between the momentum relaxation rate and the rate of ee-collisions. Hence, the interpretation of the experiments requires a subtle analysis of applicability of the approximation used.

\begin{figure}
	\centering
	\includegraphics[width=\linewidth]{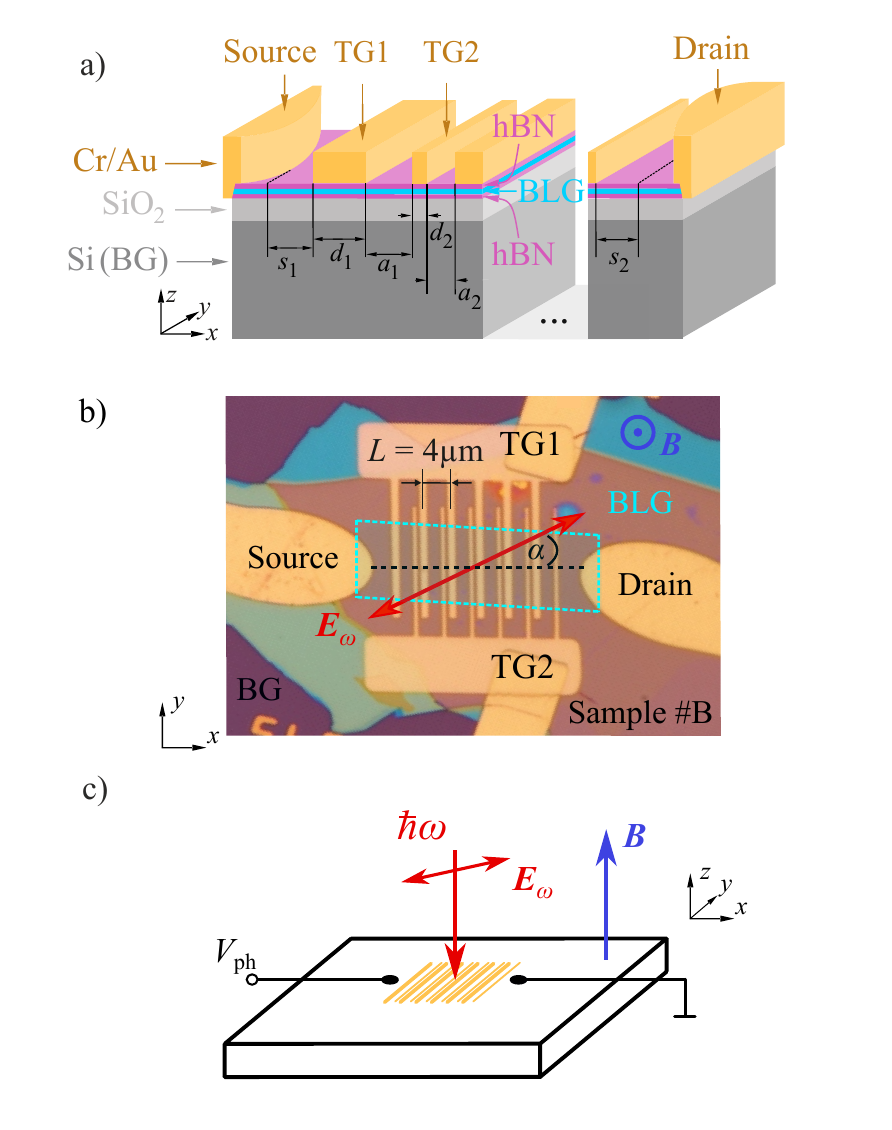}
	\caption{
		Panels~(a) and (b) depict the cross-section and a microphoto of the BLG sample with the superlattice structure fabricated on top. The former shows stacked layers of the heterostructure and assigns the parameters of the DGG structure as presented in Tab.~\ref{TabS1}. The asymmetric superlattice comprises six cells with a period $L = d_{\rm 1} + d_{\rm 2} + a_{\rm 1} + a_{\rm 2}$. Here, the widths of the metallic stripes and their spacings in between are indicated by $d_{\rm 1,2}$ and $a_{\rm 1,2}$, respectively, whereas $s_{\rm 1,2}$ labels the distance of the clean BLG flake between source/drain and the DGG. The red arrow depicts the electric field vector, $\bm E_\omega$, of the incident linearly polarized THz radiation. It's rotation, i.e it's azimuthal angle $\alpha$, is considered with respect to the orientation of the DGG stripes. Panel (c) shows the experimental setup. The photovoltage, $V_{\rm ph}$, is measured in Faraday geometry perpendicularly to the grating stripes as a photo-induced voltage drop across the sample resistance without applying external bias. 
	}
	\label{Fig1}
\end{figure}

Recently, we have demonstrated a very strong frequency dependence of the ratchet current indicating the presence of the HD regime at zero magnetic field~\cite{Moench2022}. Here we use the same samples as in Ref.~\cite{Moench2022} to study several magnetic-field- dependent effects experimentally. Using a magnetic field gives an additional way to probe the system allowing to distinguish between different regimes, and to study the effect of the ee-interaction.

\section{Samples and methods}
\label{samples-methods}

\subsection{Samples}

\begin{table*}
	\centering
	\begin{tabularx}{0.6\textwidth}{XXXX}
		\toprule[0.05cm]\addlinespace[0.2cm]
		Parameter&& 	Sample \#B 		& 	 Sample \#C \\\midrule[0.025cm]\addlinespace[0.1cm]
		flake length / width &($\SI{}{\micro\meter}$) 	&  30 / 11.5	& 17 / 6.5	\\
		top / bottom thickness of hBN& ($\SI{}{\nano\meter}$) 	&  40 / 80	& 55 / 70	\\
		\midrule[0.015cm]\addlinespace[0.1cm]
		$a_1$ / $a_2$ &($\SI{}{\micro\meter}$)	&  	2 / 0.5	& 1 / 0.25	\\
		$d_1$ / $d_2$& ($\SI{}{\micro\meter}$)	&  1 / 0.5 	& 0.5 / 0.25	\\
		$s_1$ / $s_2$& ($\SI{}{\micro\meter}$)	& 3 / 3.4 	& 2.3 / 3.1	\\
		\addlinespace[0.1cm]\bottomrule[0.05cm]
	\end{tabularx}	
	\caption{Geometric parameters of the samples \#B, and \#C. For the cross-section and top view of the structure see Fig.~\ref{Fig1}.
	}
	\label{TabS1}
\end{table*}

The samples studied in this work are high-quality BLG devices encapsulated between hexagonal boron nitrite (hBN). The structures were fabricated by using a van der Waals stacking technique \cite{Wang2013}. In addition, a Si wafer featuring a layer of 285~nm thermal SiO$_2$ on top was deployed as a substrate functioning simultaneously as a uniform back gate for the carrier-density control. In the next step, the inter-digitated DGG structures were fabricated by electron beam lithography on top of the encapsulated bilayer graphene, followed by a deposition of 5~nm Cr and 30~nm Au, and finalized by a lift-off process.
Two contacts acting as source and drain were fabricated by electron beam lithography, by subjecting the graphene layer to reactive ion etching followed by deposition of Cr and Au layers. 

Figure~\ref{Fig1} (a) and (b) show a cross-section and a microphotograph, respectively, of the sample and its inter-digitated DGG structure. The latter contains two electrically separated top gates with wide (TG1) and narrow (TG2) stripes constituting the asymmetric superlattice, with different width and spacing parameters, see Tab.~\ref{TabS1}. The superlattice comprises six cells with a period $L$. This configuration provides the possibility to bias the top gates, $U_\mathrm{TG1}$ and $U_\mathrm{TG2}$, unevenly, which enable to obtain asymmetric electrostatic potential and to control the lateral asymmetry parameter $\Xi$. 
More details on samples and their characterization can be found in Ref.~\cite{Moench2022}, where the samples with the same notations were studied.

\begin{figure}
	\centering
	\includegraphics[width=\linewidth]{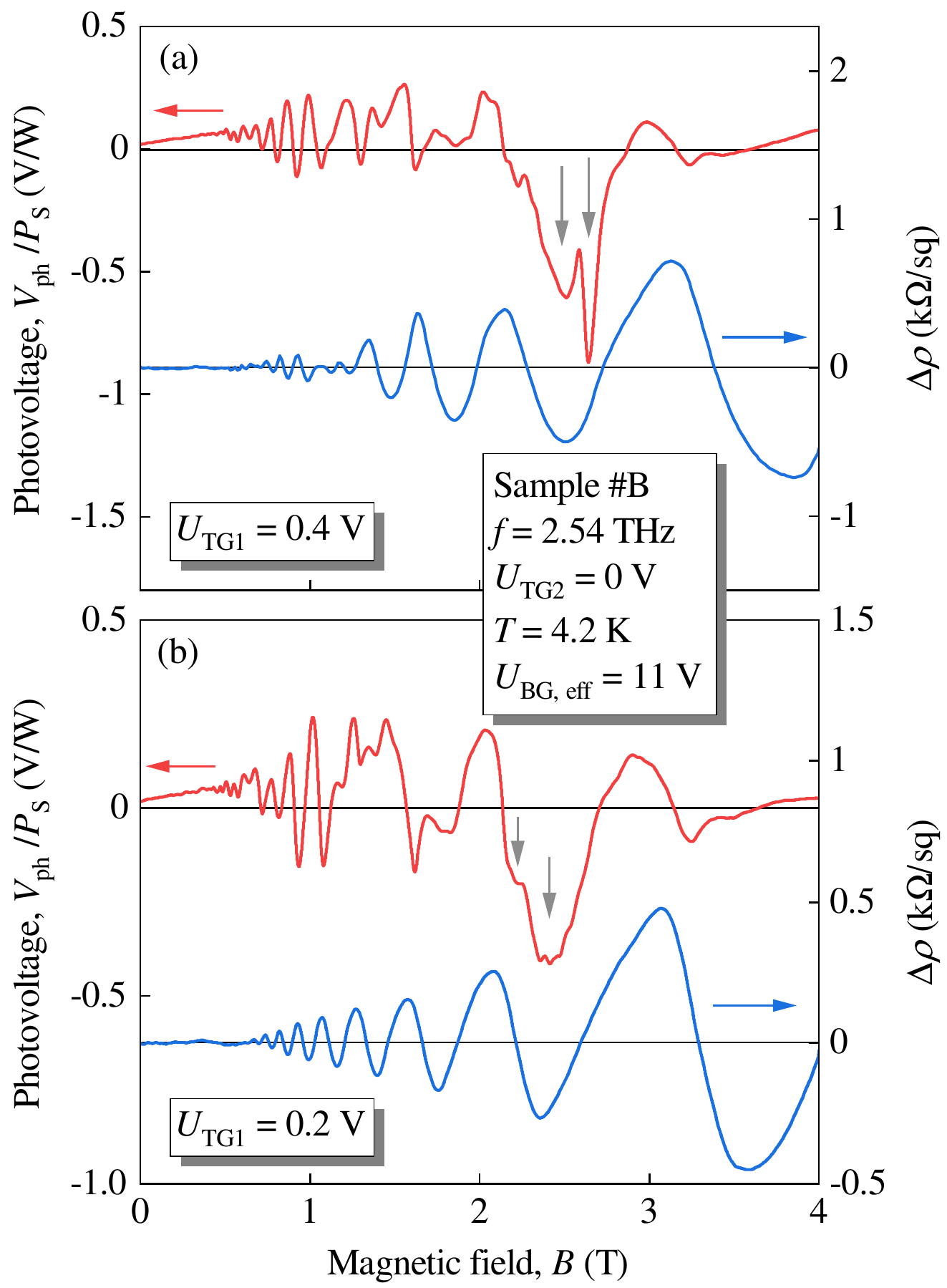}
	\caption{ Magnetic field dependencies of the photoresponse for sample \#B measured with linearly polarized radiation at a frequency $f = 2.54$~THz at high back gate voltage. Panel~(a) shows traces obtained for $U_{\rm TG1} = 0.4$~V and panel~(b) for $U_{\rm TG1} = 0.2$~V. Blue lines show corresponding dependencies of the oscillating part of sample's resistivity $\Delta\rho$ measured under the same conditions. The gray vertical arrows indicate the position of the two resonances. 	}
	\label{Fig2}
\end{figure}

\begin{figure}
	\centering
	\includegraphics[width=\linewidth]{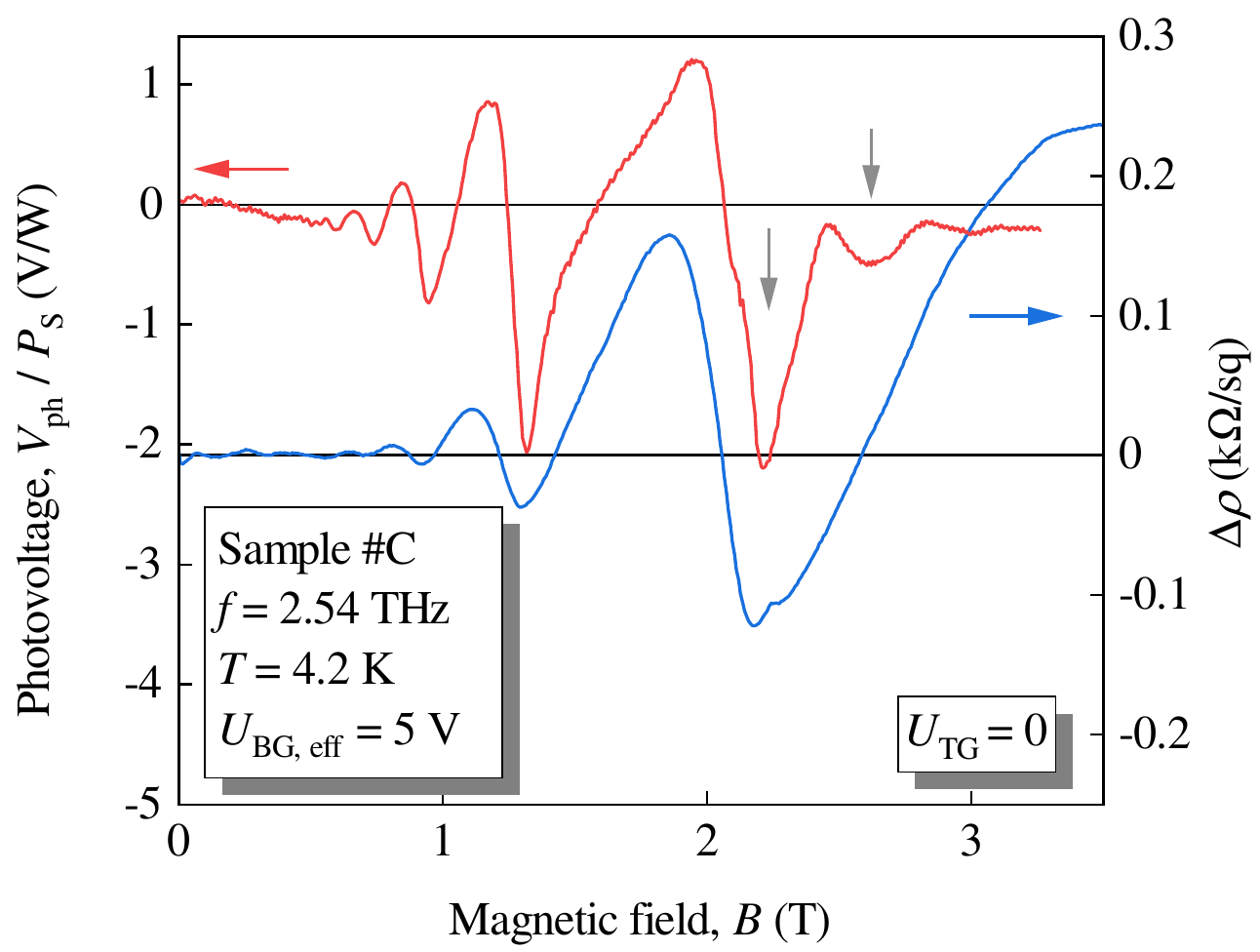}
	\caption{ Photovoltage as a function of the magnetic field measured in sample \#C with $U_{\rm BG, eff} = 5$~V and zero top gate bias. The trace was obtained for a frequency of $f = 2.54$~THz with linear polarization. Blue line depict the corresponding oscillating part of sample's resistivity $\Delta\rho$ measured under the same conditions. The position of the two resonances is marked by gray vertical arrows.
	}
	\label{Fig3}
\end{figure}

\subsection{Methods}

The ratchet photoresponse in the samples was induced by an in-plane alternating electric field $\bm{E}_\omega$. As a radiation source we used an optically pumped molecular gas laser operating in the continuous wave ($cw$) regime~\cite{Olbrich2013, Dantscher2017} with frequencies $f = 2.54, 1.63$ and 0.69~THz (corresponding photon energies are $\hbar\omega = 10.5$, 6.7 and 2.9~meV). The intensity profile was checked with a pyroelectric camera~\cite{Ganichev1999,Ziemann2000}, which showed nearly Gaussian shapes with spot diameters ranging from 1.5 to 3~mm at the sample's position. Taking into account the radiation's power $P = 80$~mW, we obtain intensities up to $I \approx 3$~Wcm$^{-2}$ (the corresponding THz electric field is $E_\omega \approx 50$~V/cm). In most experiments we used linearly polarized radiation with the electric field vector oriented with azimuth angle $\alpha$ with respect to $x$-axis i.e., perpendicular to the DGG stripes, see Fig.~\ref{Fig1}(b). The azimuth angle was varied by rotation of a $\lambda/2$ plate made of crystal quartz. In several measurements we also used right- ($\sigma^+$) and left- ($\sigma^-$) circularly polarized radiation obtained by a $\lambda/4$ wave plate with a circular polarization degree defined as $P_{\rm C}= \left[I(\sigma^+) - I(\sigma^-)\right]/\left[I(\sigma^+) + I(\sigma^-)\right]$ equals to $\pm 1$.

The samples were mounted in a temperature-controlled Oxford Cryomag optical cryostat with $z$-cut crystal quartz windows. The quartz windows were covered with black polyethylene films, which are opaque for visible- and transparent for THz-radiation. A magnetic field up to $B = 7$~T was applied in Faraday geometry and normal to the BLG plane, see \ref{Fig1}(c). The photovoltage signals $V_\text{ph}$ were picked up from  source/drain and measured using a standard lock-in technique. For that the radiation was modulated with  frequency $f_{\rm chop} = 60$~Hz using an optical chopper.
The obtained voltages were normalized to the radiation power on the sample $P_\mathrm{S} = IA_\mathrm{S}$
where $A_\mathrm{S}$ defines the area of the DGG on top of the BLG. The corresponding photocurrent $J_{\rm dc}$, generated perpendicular to the stripes of the DGG, relates to the photovoltage as $J_{\rm dc} = V_\text{ph} / R_\mathrm{s}$, where $R_{\rm s}$ is the sample resistance. 

\begin{figure}
	\centering
	\includegraphics[width=0.9\linewidth]{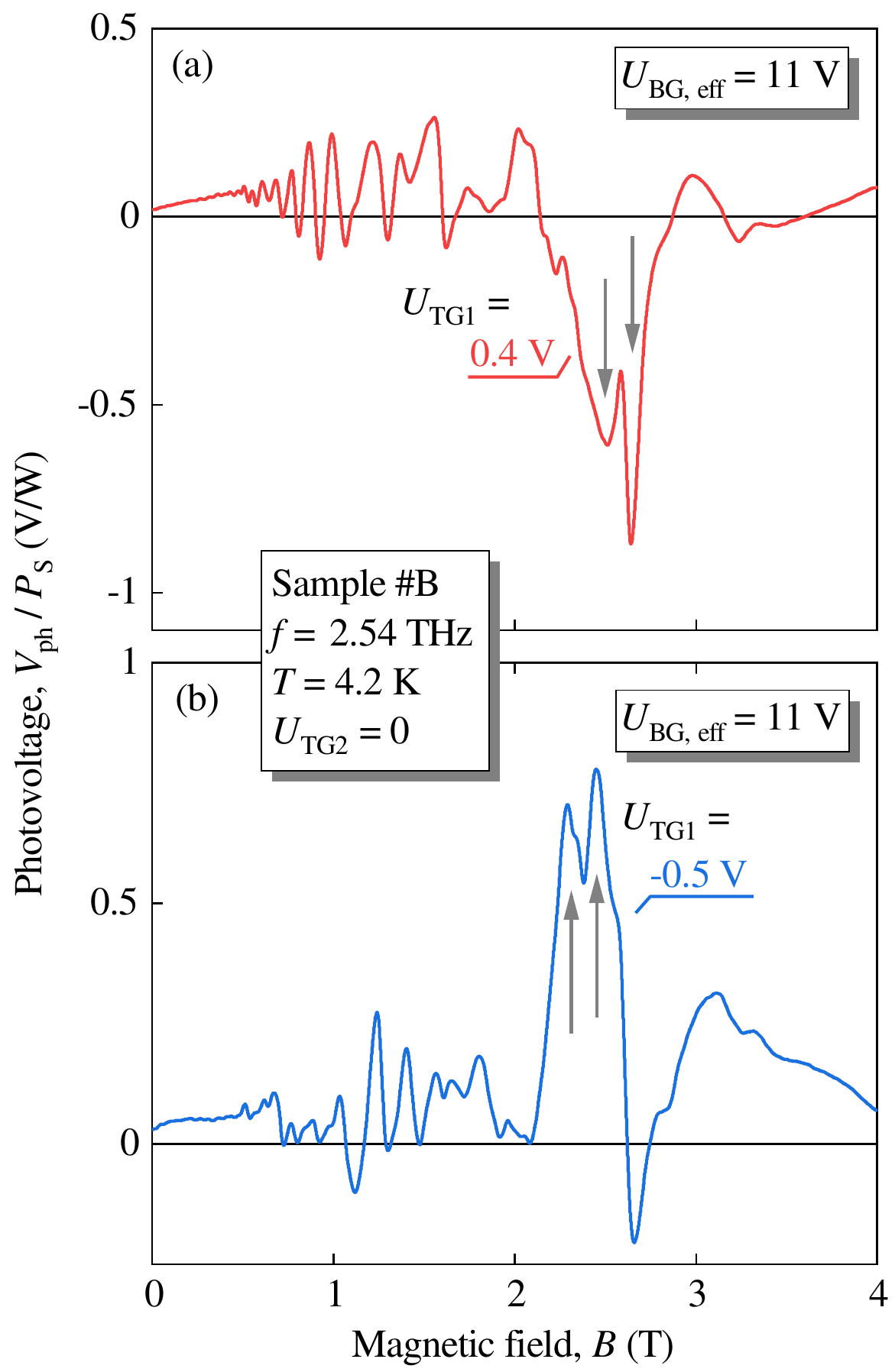}
	\caption{Magnetic field traces measured at a high back gate voltage $U_{\rm BG, eff} = 11$~V and a high TG1 voltage of opposite polarities $U_{\rm TG1} = 0.4$~V (panel a) and $U_{\rm TG1} = -0.5$~V (panel b) while holding TG2 at zero. The gray vertical arrows indicate the position of the two resonances. 
	}
	\label{Fig4}
\end{figure}

\section{Results}
\label{results}

Irradiating the DGG structure with linearly polarized THz radiation generates a photovoltage with a complex magnetic field behavior. Figures~\ref{Fig2} and \ref{Fig3} present the magnetic field dependence of the photovoltage $V_{\rm ph}$ (red curve) for samples \#B and \#C.
The data are exemplary presented for a frequency $f = 2.54$~THz and different values of back and top gate voltages. First of all, in all measurements the signal oscillates with increasing magnetic field $B$. The detected signal amplitude is more than an order of magnitude larger than the photosignal at zero magnetic field~\footnote{The zero field ratchet effect has been recently studied in details in the same structures, see Ref.~\cite{Moench2022} and, therefore, will not be discussed here.}. The oscillatory behavior of the photoresponse has also been probed for $f = 1.63$ and 0.69~THz. These measurements reveal that the amplitude strongly increases upon frequency reduction, see Appendix~\ref{frequencies}. 

Analysis of the experimental traces show that the oscillations of the photocurrent are 1/$B$-periodic and that the extremal positions of the ratchet photoresponse oscillations coincide with these of the Shubnikov-de Haas oscillations (SdHO). The oscillating part of the magnetoresistivity $\Delta\rho$ (due to a two-point measurement geometry, the Hall contribution was subtracted) is shown by blue curves in Figs.~\ref{Fig2} and \ref{Fig3}. Note that the magnetotransport was measured under the same conditions as the photovoltage.
In Appendix~\ref{SdHO_analysis} we show exemplary a 1/$B$-plot of the data for sample \#C and two different back gate voltages. We note that while the amplitude of the SdHO of the resistivity as well as the amplitude of the photoresponse oscillations in sample \#C increases with rising magnetic fields, in  sample \#B, we observed a non-monotonic behavior at $B \approx 1$~T. Figure~\ref{Fig2}(b) shows that the amplitude of the oscillations first increases, exhibits a maximum and then decreases with increasing magnetic field.

Strikingly, for magnetic fields around $B \approx$ 2.5~T the oscillations are superimposed by two resonances dominating the photoresponse, see gray arrows in Figs.~\ref{Fig2} and~\ref{Fig3}. These experimental traces will also be compared to the developed theory below, which allows to identify the origin of the resonances. For large back gate voltages and with applied top gate bias the observed resonances are particularly narrow and sharp with a width less than half a period of the SdHO. Consequently, they can be studied without much influence of the oscillating part of the photoresponse. We now focus on the results obtained under similar conditions.

Figure~\ref{Fig4}(a) and (b) show a close-up view of the resonances measured for opposite polarity of the top gate bias $U_\mathrm{TG1}$. Importantly, reversing the polarity results in a sign change of both resonant photoresponses, and consequently, inverses the photocurrent direction. This observation proves clearly that the photocurrent is caused by the ratchet effect. Indeed, the sign of the ratchet photocurrent is defined by the sign of the lateral asymmetry parameter $\Xi$, see Eq.~\eqref{Xi0}, which reverses upon inverting the top gate polarity. Note that the link between the photocurrent and the lateral asymmetry parameter is also demonstrated for the oscillatory part of the signal, see Appendix~\ref{top_gate_dep}.

\begin{figure}
	\centering
	\includegraphics[width=\linewidth]{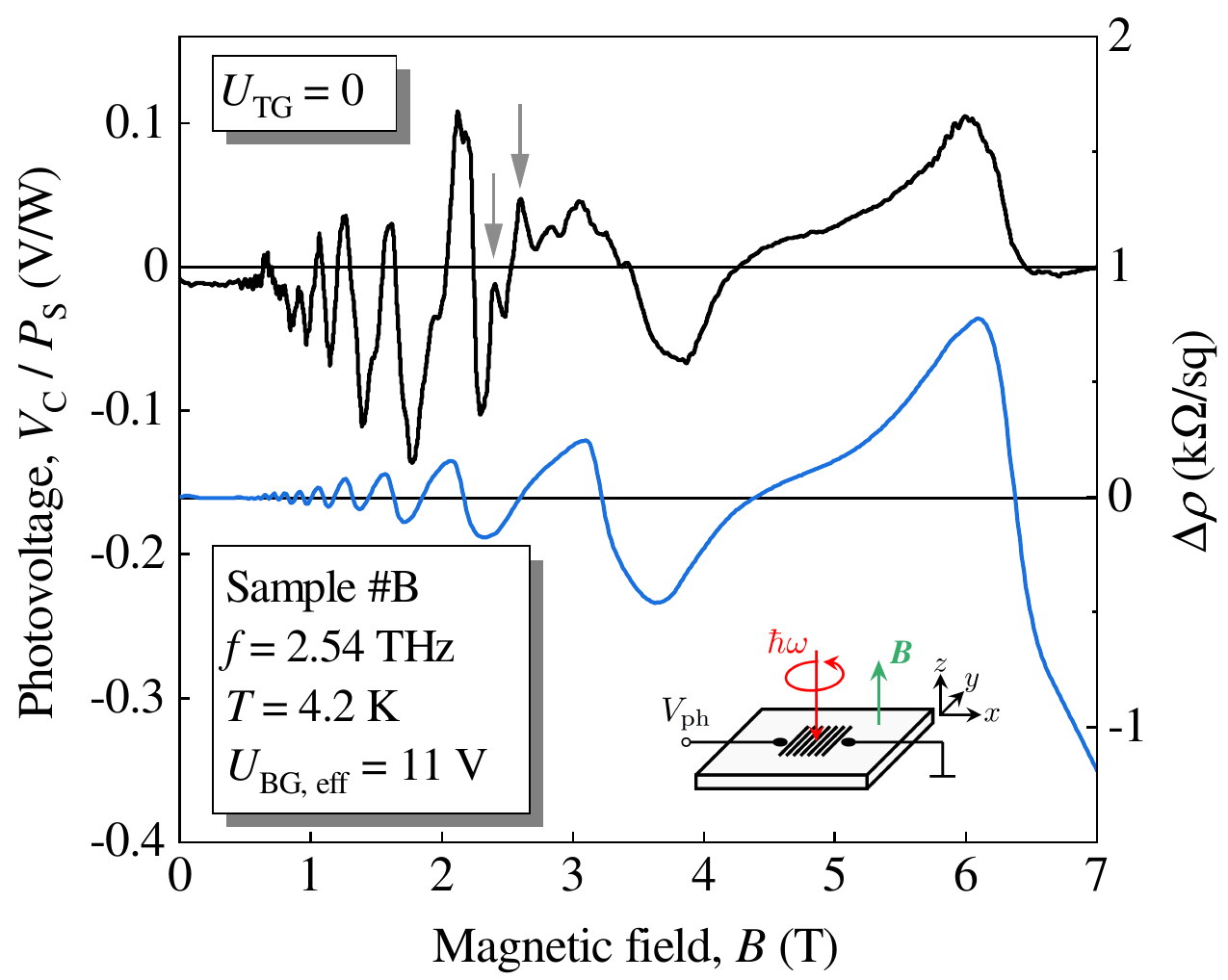}
	\caption{
	Magnetic field trace of the circular contribution to the magneto-ratchet effect measured in sample \#B at $f = 2.54$~THz. The curve is obtained by measuring the photoresponse for right- ($\sigma^+$) and left-handed ($\sigma^-$) circularly polarized radiation and calculating the circular contribution after $V_\mathrm{C} = \left[V_\mathrm{ph}(\sigma^+) - V_\mathrm{ph}(\sigma^-)\right]/2$. The inset shows the measurement configuration. 
	The gray vertical arrows mark the position of two resonances.}
	\label{Fig5}
\end{figure}

 The magneto-ratchet photoresponse described so far was obtained for linearly polarized radiation. Varying the relative orientation of the radiation electric field vector $\bm E$ with respect to the DGG stripes, we observed that the photoresponse varies after $V_{\rm ph} = V_0 + V_{\rm L}\cos(2\alpha + \theta)$, with a phase angle $\theta$, a polarization independent $V_0$ and a polarization dependent contribution $V_{\rm L}$, which is several times larger than $V_0$. We also performed measurements applying right- and left-handed circularly polarized radiation. Figure~\ref{Fig5} reveals that the excitation with THz radiation also results in a helicity sensitive photocurrent $J_{\rm C} \propto V_\mathrm{C} \propto P_{C}$, which is calculated according to $V_\mathrm{C} = \left[V_\mathrm{ph}(\sigma^+) - V_\mathrm{ph}(\sigma^-)\right]/2$. While the amplitude of the helicity-driven photocurrent is substantially smaller than $V_{\rm ph}$, the overall qualitative behavior remains the same: the photosignal exibits sign-alternating oscillations with the amplitude much larger than the signal at zero fields, and two resonances appear in the vicinity of $B = 2.5$~T, see gray  arrows in Fig.~\ref{Fig5}.

\section{Theoretical model and approaches}
\label{model}

Before we develop a theory of the magneto-ratchet effect under study we introduce a basic model.
We consider the  motion of 2D  electrons with parabolic spectrum
in the external field $\bm E_\omega$ of general polarization described by phases $\alpha$ and $\theta$:
\be
E_{\omega,x} = {E_{0}} \cos \alpha \cos {\omega}t, \quad
E_{\omega,y} = E_{0}\sin \alpha \cos \left( {{\omega}t + \theta } \right).
\label{Eext}
\ee
The phases $\alpha$ and $\theta$
phases are
connected with the standard Stokes parameters (normalized by $E_0^2$) as follows:
\be
\begin{aligned}
	& P_0=1,  \quad P_{\rm L1}=\cos(2\alpha),
	\\
	& P_{\rm L2}=\sin(2 \alpha) \cos \theta, \quad  P_{\rm C}=\sin(2\alpha) \sin \theta.
	\label{Stockes}
\end{aligned}
\ee

We assume that  the static  periodic  potential
\be
\label{V}
V(x)   = V_0\cos qx,
\ee
($q=2\pi/L$) arises in the 2D gas due to presence of the grating gate structure with the period $L$.
We also assume that the grating leads to modulation of the electric field with the depth $h,$
so that the field acting in the 2D channel,
\begin{equation}
\label{E_x_t}
\bm E(x,t) = \bm E_0(x) \text{e}^{-i\omega t} + \bm E_0^*(x) \text{e}^{i\omega t},
\end{equation}
has spatially modulated  complex amplitude $\bm E_0(x)$  with the components
$E_{0x}(x)= E_0 \cos \alpha [1+ h \cos(q x+\varphi)]/2,$
$E_{0y}(x)= E_0 \sin \alpha  e^{-i\theta} [1+ h \cos(q x+\varphi)]/2,$
where $\varphi$ is the phase which determines the asymmetry of the
modulation.
The components of the total field in the channel read
\begin{align}
\label{Ex}
E_x (x,t) & = [1+ h \cos(q x+\varphi)]  E_0 \cos \alpha  \cos \omega t , \nonumber
\\
E_{y}(x,t) & = [1+ h \cos(q x+\varphi)]  E_0 \sin \alpha  \cos (\omega t+\theta).
\end{align}

We search the response of the  2D electron system to the above described perturbation
by using two approaches --- hydrodynamic  and drift-diffusion.
In both approaches,  the direction of the current is determined by the
frequency- and disorder-independent
asymmetry parameter
\begin{equation}
\label{Xi}
\Xi= \overline{ {dV\over dx}|\bm E_0(x)|^2}= \frac{E_0^2 h V_0 q \sin \varphi  }{4} ,
\end{equation}
which  oscillates with changing of the phase shift $\varphi.$
HD and DD regimes  are realized depending on relation between the electron-electron and electron-impurity scattering times.

For describing  DD case, we neglect ee-interaction and use the Boltzmann equation  where the  impurity scattering rate  $\gamma=\gamma(E_{\rm F})$ contains   the correction rapidly oscillating  with the Fermi energy, $E_{\rm F}$.  In the HD case, we use hydrodynamic equations  with $\gamma=\gamma(N_0),$ where  $\gamma(N_0)$ rapidly oscillates with the electron concentration  $N_0$. In addition to fast ee-collisions,  within HD approach, we  take into
account the long-range interaction  leading to plasmonic effects.      In both cases, we  assume that the main contribution to the ratchet current comes from the second derivative of  rapidly oscillating scattering rate:    $j_{\rm DD}\propto {\dd^2 \gamma}/{\dd E_{\rm F}^2}$ and        $j_{\rm HD}\propto {\dd^2 \gamma}/{\dd  N_0^2}$.
Both  DD and HD approximations  predict CR resonance at $\omega=\omega_c$. This resonance is clearly  seen in the experiment.  We also see a splitting of the resonance onto two sub-peaks. We interpret this splitting  as appearance  of   a MP satellite at
\be
\omega=\sqrt{\omega_c^2+ \omega_q^2},
\label{MP}
\ee
where $\omega_q=s q $ is  plasmonic frequency corresponding to density modulation with the wave vector, $q,$  of lateral superlattice, and $s$ is the plasma wave velocity.  Remarkably, this is not a simple plasmonic shift of CR  but a   coexistence of
CR resonance   and
its plasmonic satellite.
We describe the splitting  within HD approximation accounting for long-range interaction.

 One of the  most interesting  theoretical results obtained within  DD approximation  is the resonance  at the  second harmonic of CR, at $\omega=2 \omega_c$.  Remarkably, in the absence of  ee-collisions this resonance  is parametrically larger  than CR, by a factor $\sim \omega_{\rm c} \tau_{\rm tr} \gg 1$, where $\tau_{\rm tr}$ is the transport scattering time determined the mobility. Also, its shape is different as compared to CR: The second-harmonic   resonance is asymmetric and is described by a derivative of the Lorentz peak. With increasing  $\gamma_{\rm ee}$ above $\gamma$ the resonance at the   second harmonic  broadens and simultaneously  decreases in amplitude. In the ideal liquid  limit,  $\gamma_{\rm ee} \to \infty,$ the second harmonic must disappear. Hence, we have an  additional tool for distinguishing between DD regime, where second harmonic should dominate over CR, and   HD regime, where resonance at $\omega=2\omega_{\rm c}$  is strongly suppressed. Below, we try to fit our data by the expressions obtained within both DD and HD approaches.

\color{black}

\subsection{Drift-diffusion approach: no ee-interaction}

We use the Boltzmann kinetic equation approach, where the electric
current density is given by the following expression
\begin{equation}
\label{j}
\bm j = e \sum_{\nu, \bm p} \bm v_{\bm p} \bar{f}_{\bm p}.
\end{equation}
Here $\bm v_{\bm p} = \bm p/m$ is the velocity of carriers having the momentum $\bm p$ with $m$ being the effective mass, $\nu$ enumerates spin and valley-degenerate states, and $\bar{f}_{\bm p}$ is the  distribution function $f_{\bm p}(x)$ averaged over the space period of DGG structure. The latter is a solution of the kinetic equation~\cite{Budkin2016a}
\begin{equation}
\label{kin_eq}
\left( {\partial \over \partial t} + v_{\bm p, x}{\partial \over \partial x} + \bm F_{\bm p} \cdot {\partial \over \partial \bm p} \right) f_{\bm p}(x) = \text{St}[f_{\bm p}(x)].
\end{equation}
Here $\text{St}[f]$ is the elastic scattering collision integral, and the space- and time-dependent force is given by
\begin{equation}
\bm F_{\bm p} = e\bm E(x,t) + {e\over c}\bm v_{\bm p}\times \bm B - {\dd V\over \dd x} \hat{\bm x},
\end{equation}
where $\bm E(x,t)$ is given by Eq.~\eqref{E_x_t}.
The distribution function is found by sequential iterations of the
kinetic equation in small electric field amplitude and the ratchet
potential with the result linear in $\dd V/\dd x$ and quadratic in  $|\bm
E_0(x)|$ with the ratchet current proportional to the asymmetry
parameter $\Xi$ given by Eqs.~\eqref{Xi0},~\eqref{Xi}.

The  calculation are quite standard and fully analogous to the ones presented in~Ref.~\cite{Hubmann2020}.  General expression for  the ratchet current density is given by
\be
\begin{aligned}
	\label{j_tot}
	&j_x+ij_y = j_0
	\delta_c\left( {2\pi E_\text{F}\over \hbar\omega_c}  \right)^2
	\\
	& \times  (P_0 D_0 + P_{L1} D_{L1} + P_{L2} D_{L2} + P_C D_C).
\end{aligned}
\ee
Here $j_0$ is the linear ratchet current density in the absence of magnetic field and at $\omega\tau_{\rm tr} \ll 1$:
\begin{equation}
\label{j0}
j_0 =\Xi {e^3  \tau_{\rm tr}^3\over \pi \hbar^2 m},
\end{equation}
and the dimensionless amplitudes  $D_0$,  $D_{L1}$,  $D_{L2}$,  and $D_C$ depend on the light frequency, magnetic field and impurity scattering strength via the parameters $\omega/\omega_c$ and $\omega\tau_{\rm tr}$. Analytical expressions  for the amplitudes  are quite cumbersome and presented in the Appendix~\ref{Supplemental} [see Eq.~\eqref{D_i_DD}].
The oscillations are encoded in the  coefficient
\begin{equation}
\label{delta_c}
\delta_c=2\cos{(2\pi E_\text{F}/\hbar\omega_c)}\exp(-\pi/\omega_c\tau_q){\chi\over \sinh\chi},
\end{equation}	
where $\chi= 2\pi^2 k_\text{B}T/\hbar \omega_c$, and $\tau_q$ is the quantum relaxation time.
The second derivative of $\delta_c$,
\be
\label{Eq15}
E_F^2\delta_c''= -\left( {2\pi E_\text{F}\over \hbar\omega_c}  \right)^2 \delta_c
\ee
gives factor  $\left( {2\pi E_\text{F}/ \hbar\omega_c}  \right)^2 ,$ which is responsible for giant enhancement of the photoresponse in the SdH regime.

\subsection{Hydrodynamic approach: strong ee-interaction}
Hydrodynamic   equations  for concentration  and velocity   read
\begin{align}
\label{n}
& \frac{\p n}{ \p t}+ {\rm div}~[(1+n) \m v]=0, \\
\label{v}
& \frac{\p \m v}{ \p t}+ (\m v {\m \nabla}) \m v+  \gamma (n) \m v +\boldsymbol{ \omega}_c\times \m v + s^2 {\m \nabla} n =\m a.
\end{align}
Here
\be
n= \frac{N -N_0}{N_0},
\ee
$N=N(x,t)$  is the concentration in the channel and $N_0$ its equilibrium value, $\m v$ is the drift velocity,
\be
\m a=  \frac{e \m E}{m} - \frac{{\m \nabla} V}{m}  ,\quad
\m E=
\left(
\begin{array}{c}
	E_x \\
	E_y \\
\end{array}
\right),
\ee
$\boldsymbol{\omega}_c = \omega_c \hat{\bm z}$  is the cyclotron angular velocity, $s$ is the  plasma waves velocity, and
$\gamma(n)\equiv 1/\tau_{\rm tr}$ is the momentum relaxation rate.
We neglect in  Eq.~\eqref{v} a viscous term $\eta \Delta \m v,$ assuming that  $\eta q^2 \ll 1/\tau_{\rm tr}$  (here $\eta $ is the electron viscosity).  

Direct current appears due to  non-linearity caused   rectification provided that  structure is asymmetric, i.e.   $\Xi \neq 0$ (see Eq.~\eqref{Xi}).
The non-linearity is encoded  in several terms. First of all, there are standard  non-linear hydrodynamic terms
$\p (n \m v)/\p x,$ $(\m v \boldsymbol{\nabla})\m v  .$ What is more important,  the
transport scattering rate
$\gamma$   depends on local Fermi energy and, consequently, on the dimensionless concentration $n.$  At zero magnetic field this dependence
is irrelevant but plays a key role in the regime of SdH oscillations where $\gamma$ contains small but
rapidly oscillating correction. We assume that  response is determined by non-linear  term $\gamma''(0) n^2 \m v/2$ in Eq.~\eqref{v} which arises after expansion $\gamma(n)$ over $n$ up to the second order.

The responsivity in the regime of SdH oscillations
increases not only in grating gate structures but also in single
field-effect transistors (FETs)~\cite{Sakowicz2008,Lifshits2009,BoubangaTombet2009,Klimenko2010}.
Although  the general physics of enhancement in both cases is
connected with fast oscillations  of the resistivity, there is an
essential difference. In grating gate structures the shape of typical
\textit{dc} photoresponse roughly reproduces  resistance oscillations, while
in single FETs the typical response is $\pi/2$ phase shifted with respect
to resistance oscillations. The latter shift was explained
theoretically by a hydrodynamic model in Ref.~\cite{Lifshits2009} and demonstrated experimentally in Ref.~\cite{Klimenko2010}. The key idea is as follows. Transport scattering rate
$\gamma(n) $ in the SdH regime sharply depends on the dimensionless
electron concentration $n$.
Expanding $\gamma(n)\approx \gamma(0)~ +~ \gamma'(0)
n$ with respect to small $n,$ one finds that a nonlinear term,
$\gamma'(0) n \m v$,   appears in the Navier-Stokes Eq.~\eqref{v}. This is sufficient to give a nonzero
response, which  in a single FET arises in the second order with
respect to external THz field  	(both $n$ and $\m v$ are linear with
respect to this field). Therefore, in this case, the response is
proportional 	to  the first derivative of $\gamma' (0)$ with respect to
concentration,
hence, it is
$\pi/2$ shifted in respect to the conductivity oscillations. By
contrast, in the  grating gate structures, the \textit{dc} response appears
only in the third order with respect to the perturbation \cite{Ivchenko2011}.
As a consequence, the ratchet current is proportional to the second
derivative $\gamma''(0)$ (see discussion in Ref.~\cite{Sai2021} and  below  in  Appendix~\ref{AppHD}) and therefore
roughly (up to a smooth envelope) reproduces resistance oscillations.

In order to solve HD equations, we
use approximation  suggested
in Ref.~\cite{Lifshits2009} for analysis of response of a single FET.  We assume that  $\gamma=\gamma(n)$ is the oscillating function
of $n$  because of the SdHO
\be
\gamma(n)=\gamma_0\left(\!  1- 2 \delta_c^{\rm HD} \right).
\ee
Here
\be
\delta_c^{\rm HD}=2\frac{ \chi}{\sinh \chi} \exp\!\!\left(\!- \frac{\pi}{\omega_c\tau_q}
\!\right) 
\cos\!\! \left(\frac{2\pi E_{\rm F}(x,t)}{\hbar \omega_c}\right)\!,
\ee
where 
$E_{\rm F}(x,t)=E_{\rm F}^0[1+n(x,t)]$
is related to the concentration in the channel as $N(x,t)=g E_{\rm F}(x,t)$
with $g$ being the density of states,  and $\tau_q$ is quantum scattering time, which can be strongly renormalized by electron-electron collisions in the hydrodynamic regime.
We expand this function near the Fermi level:
\be
\gamma(n) = \gamma(0) + \gamma'(0) n  + \gamma''(0) \frac{n^2}{2},
\label{gamma-n}
\ee
where  $\gamma'$   and   $\gamma''$ are, respectively, first and second derivatives  with respect to $n$ taken  at the Fermi level.       Since  the oscillations are very fast, we assume
\be
\frac{\gamma'}{\gamma} \sim \frac{\gamma''}{\gamma'} \sim \frac{ E_F}{\hbar \omega_c} \gg 1.
\label{condition}
\ee
Due to these inequalities the oscillating contribution to
the ratchet current can be very large, and its amplitude substantially exceeds the zero-field value~\cite{Hubmann2020}.
Here, we focus on oscillations of the ratchet effect, so that we only
keep oscillations  related to dependence of $\gamma$ on $n.$

We use  the  same  method of calculation  as one developed in Ref.~\cite{Rozhansky2015}
for hydrodynamic analysis  of the ratchet effect at zero magnetic field.
Specifically,
we  use perturbative   expansion of $n$ and $\m v$  and  dc  current,
\be
\bm j = -eN_0
\left \langle (1+n)\m v \right \rangle_{t,x}\ee  over
$E_0$ and $V.$  Non-zero contribution, $\propto E_0^2 V_0$,
arises   in the 2nd order in $E_0$ and in the first order in $V_0$  [see Eq.~\eqref{Xi}].
Let us formulate the key steps of calculations.
As we explained above, due to the  large parameter, ${ E_F}/{\hbar \omega_c} \gg 1,  $   the main contribution
to the  rectified ratchet current comes from the non-linear term $\gamma''(0) \m v  {n^2}/{2}$
in the r.h.s. of Eq.~\eqref{v} [see also Eq.~\eqref{gamma-n}]. We therefore  neglect all other nonlinear
terms in the hydrodynamic equations.  Calculating $n$ and $\m v$ in linear (with respect to
$E_0$ and $V$) approximation, substituting  the result into non-linear term and
averaging over time and coordinate, we get
$\gamma''(0)\langle  \m v  {n^2}\rangle_{x,t}/2 \neq 0. $  Next, one can
find rectified current  by   averaging of Eq.~\eqref{v} over $t$ and $x. $ This procedure
is a quite standard.  
The result is given in Appendix~\ref{AppHD} (see also Ref.~\cite{Sai2021} for details of the calculations).
The ratchet current can be also presented in the same form as the DD result [cf. Eq.~\eqref{j_tot}]
\be
\begin{aligned}
	\label{j_tot_HD}
	&j_x+ij_y = \tilde{j}_0
	\delta_c\left( {2\pi E_\text{F}\over \hbar\omega_c}  \right)^2
	\\
	& \times  (P_0 \tilde D_0 + P_{L1} \tilde D_{L1} + P_{L2} \tilde D_{L2} + P_C \tilde D_C).
\end{aligned}
\ee
Here 
\begin{equation} 
\tilde j_0= j_0 
\frac{v_{\rm F}^2}{2 s^2}
\label{tildej0}
\end{equation} 
[$j_0$ is given by  Eq.~\eqref{j0}], and  dimensionless amplitudes $\tilde D_{0, L1, L2, C}$ are presented in Appendix \ref{AppHD} [see Eq.~\eqref{tildeD }].
A  special comment is needed about factor $v_{\rm F}^2/2 s^2.$ This factor appears due to the screening effect, which is taken into account in HD approach but missed within DD approximation that fully neglects ee-interaction. One can show that  in the absence of interaction this factor turns to unity. We expect that including  long-range interaction  into DD calculations would also lead to the same screening effect.

\section{Theoretical results on the resonant magnetic-ratchet effects}
\label{theory_results}

General formulas for the ratchet current oscillation amplitudes are quite cumbersome both in HD and DD case, so that we present them in Appendix~\ref{Supplemental}. Here we present results for both regimes in the simplest limiting cases having in mind to focus on comparison with experiment and on discussing differences between HD and DD approximations.

\subsection{Drift-diffusion}

First we focus on the CR resonance where, according to our measurements, the ratchet current is strongly enhanced.
In agreement with the experimental results, the DD theory  yields a resonance enhancement of the ratchet current oscillation amplitude in vicinity of the main CR harmonics $\abs{\omega_c} \approx \omega$.  From Appendix~\ref{Supplemental}, Eq.~\eqref{DD_resonance}, we obtain that the ratchet current is given by
\begin{equation}
\label{DD_res_CR}
j(\omega \approx \pm {\omega_c}) =
{j_0 \over (\omega\tau_{\rm tr})^3}\qty({2\pi E_\text{F}\over  \hbar \omega})^2 \delta_c
{P_\text{\rm C} \pm 1\over 1+[(\omega-\abs{\omega_c})\tau_\text{tr}]^2},
\end{equation}
where $j_0$ is given by Eq.~\eqref{j0}.
Note that the $y$-component dominates in the ratchet current in this approximation.

Now we turn to the resonance $\omega = \pm 2\omega_c$. In this case, both components of the ratchet current are present, see Eq.~\eqref{DD_2_omega_c_res}. The $x$- and $y$-components depend on frequency as an antisymmetric and symmetric contours, respectively:
\begin{equation}
\label{CR2}
j_x+ij_y= -
{j_0 \over (\omega\tau_{\rm tr})^3}{\tau_2\over \tau_\text{tr}}\qty({2\pi E_\text{F}\over  \hbar \omega})^2
\delta_c
{1\pm P_\text{\rm C} \over 1+\epsilon_2^2} (\epsilon_2 \pm i).
\end{equation}
Here
\[
\epsilon_2= (\omega\mp 2\omega_c)\tau_2
\]
with $\tau_2$ being the relaxation time of the second angular harmonics of the distribution function, see Appendix~\ref{Supplemental} for details.

It is important that the relaxation time $\tau_2$ is governed by both the impurity and ee-scatterings while the transport time $\tau_\text{tr}$ is immune to the ee-interaction strength. Therefore the amplitude of the second CR harmonics being $\propto \tau_2/\tau_\text{tr}$ is determined by the interplay of the ee- and impurity-scattering strengths. In particular, it follows from Eq.~\eqref{CR2} that the second CR harmonics amplitude  is suppressed if the ee-scattering is strong.

\subsection{Hydrodynamics }

Importantly, the ratchet current  encodes information both about cyclotron
resonance, which happens at  ${\omega=  |\omega_{\rm c}|}$  and about MP resonance,
which happens at frequency given by Eq.~\eqref{MP}.
Physics behind two peaks is as follows. Oscillating electric field contains two components: homogeneous oscillating field, $E_1 \propto \cos (\omega t),$ and modulated contribution, $E_2\propto \cos(qx+ \varphi) \cos (\omega t).$ First contribution corresponds to $q=0$ and, consequently,    excites CR resonance  exactly at the cyclotron frequency.  The second contribution leads to MP resonance. Since non-zero response appears in the third order with respect to external field, $j_{\rm dc } \propto E_1 E_2 \dd V/\dd x $ (here $V$ is the static potential), it should show both CR and MP resonances simultaneously. The interference of $E_1$ and $E_2$ contributions   leads to Fano-like asymmetry of the resonances [see Eqs.~\eqref{resonance} and \eqref{passive} below] \cite{note,Potashin2020}. 

Let us describe CR  and MP resonances on equal footing within the resonance  approximation.
To this end, we consider an envelope of rapidly oscillating current given by Eq.~\eqref{supp-J} and simplify it under the resonance conditions, $\abs{\omega-\abs{\omega_{\rm c}}}, \gamma \ll \omega$.
We also assume that ${\omega_q \ll \abs{\omega_{\rm c}}}$,  so that MP resonance occurs at
$\omega\approx\omega_{\rm c}  + \omega_q^2 /2\omega_{\rm c}$ (here, we assume $\omega_{\rm c} >0$). The CR and MP  resonances  become
distinguishable provided that $\omega_q^2 /2\omega_{\rm c}$ becomes  on the order of the
resonance width $\gamma.$  Next,  we introduce dimensionless parameters
\begin{equation}
\epsilon=\frac{\omega-\omega_{\rm c}}{\gamma},
\qquad
\Delta = \frac{\omega_q^2}{2\omega_{\rm c} \gamma},
\label{Delta}
\end{equation}
 and keep in the envelope of  Eq.~\eqref{supp-J}  a leading non-zero term with respect  to  the small parameter $\gamma/\omega_{\rm c}$ (for fixed  $\epsilon$ and $\Delta$). Finally, we multiply  thus found  envelope  on oscillating factor $\delta_c$ and get resonance expression for the current, which   depends on the polarization.  Interestingly, results  for linear polarization along the $x$ axis ($P_0=P_{\rm L1}=1,~P_{\rm C}=P_{\rm L2}=0$) and  for the circular active polarization   ($P_0=-P_{\rm C}= 1,~P_{\rm L1}=P_{\rm L2}=0$) coincide
 \begin{align}
 j_x^{\rm lin}
 =j_x^{\rm circ,act}
 &= \tilde j_0\delta_c \left( \frac{2\pi E_F}{\hbar \omega_c}\right)^2 \nonumber
 \\
 & \times \frac{2\gamma^2(1-\epsilon  \Delta+\epsilon^2)}{\omega_{c}^{2}(1+\epsilon^2)\left[1+ (\epsilon -\Delta)^2 \right]}. 
 \label{resonance}
 \end{align}
 Here $\tilde j_0 $  and $\delta_{\rm c}$ are given by Eq.~\eqref{tildej0} and  Eq.~\eqref{delta_c}, respectively.
 It worth noting that, for the same circular polarization, there is also parametrically smaller (by a factor $\sim \gamma^2/\omega_{\rm c}^2$)   MP resonance  of opposite sign for $\omega \approx -\omega_{\rm c }- \omega_q^2 /2\omega_{\rm c}$ (passive resonance)
 \begin{equation}
 j_x^{\rm circ,pass}
 = -\tilde j_0 \delta_c \left( \frac{2\pi E_F}{\hbar \omega_c}\right)^2
 \frac{\gamma^4(1+ \tilde \epsilon  \Delta+\tilde \epsilon^2)}{\omega_{c}^{4}\left[1+ (\tilde \epsilon +\Delta)^2 \right]}.
 \label{passive}
 \end{equation}
 Here $\tilde \epsilon=(\omega+\omega_c)/\gamma.$

Equation~\eqref{resonance} clearly shows a splitting of CR and MP resonances. Indeed, in the absence  of ee-interaction, when $\Delta=0,$ we have  the Lorentz peak $\propto 1/(1+\epsilon^2)$ corresponding to CR. With increasing $\Delta$ this peak splits onto  CR at $\epsilon=0$ and MP resonance at $\epsilon=\Delta.$ We notice that for $\Delta \gg 1  ,$    these resonances are well separated and have interference-induced asymmetric shape given by the first derivative of the Lorentz peak.
 For passive circular polarization, only MP is present as follows from  Eq.~\eqref{passive}.  In particular, at $\Delta=0$ this resonance disappears.

\subsection {Frequency dependence of the response: Comparison of DD and HD regimes}

\begin{figure}
	\centering
	\includegraphics[width=0.42\textwidth]{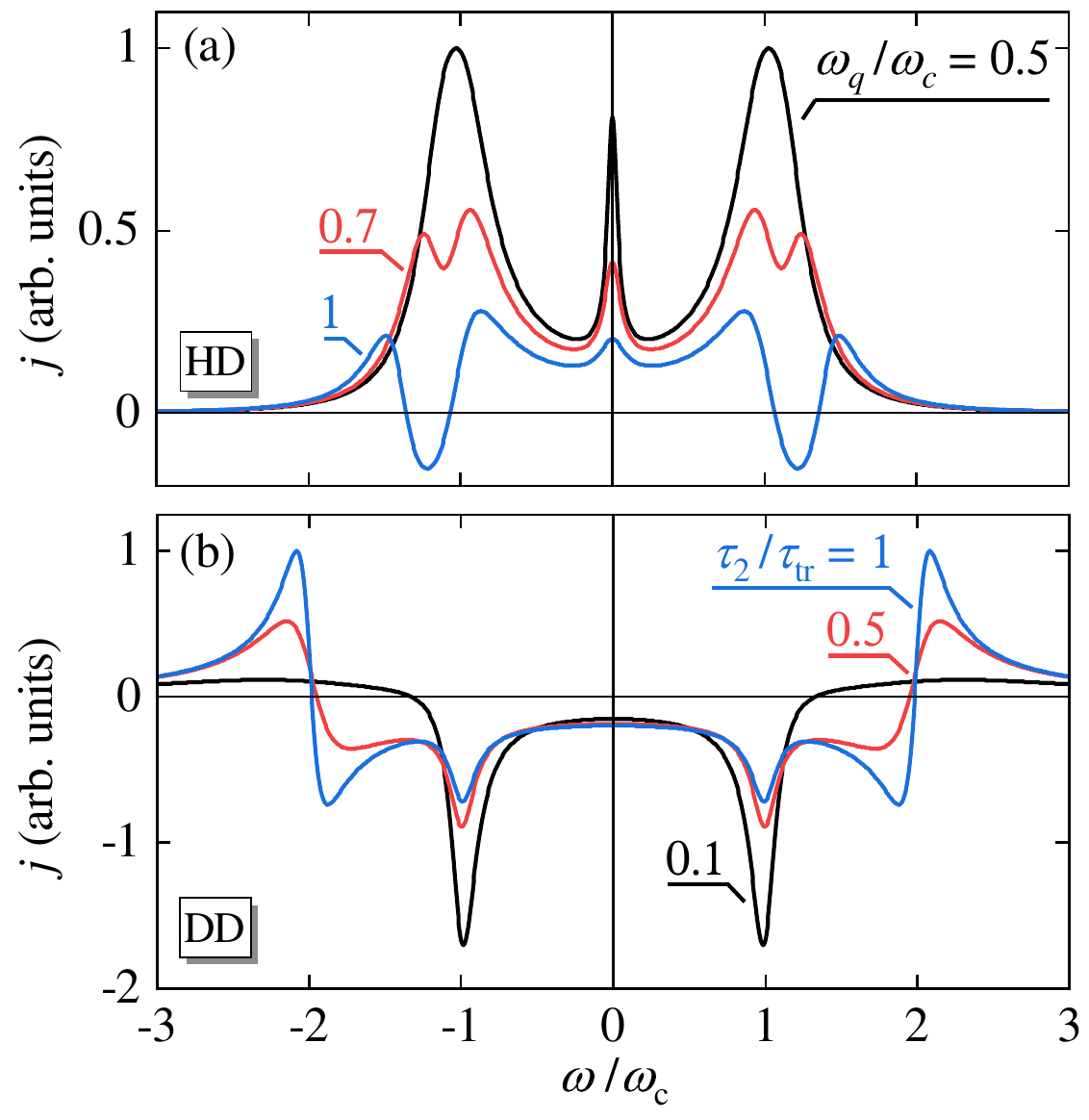}
	\caption{  Current  $j$  in frequency-independent  
 units  (see definition of these units in the text)   as a function of $\omega/\omega_{c}$  in the HD regime (a) and  in the  DD regime (b). Here the ratio $\gamma/\omega_{\rm c}$ is fixed by the value  $0.2.$ All curves are calculated for linear polarization directed along the $x$-axis.  In panel (a) we illustrate  the  evolution of current with changing the ``plasmonic parameter''  $\omega_{q}/\omega_{c}.$    Curves in panel (b) are plotted for different values of the ratio $\tau_{2}/\tau_{\rm tr}$. }
	\label{Fig6}
\end{figure}

In our experiments, we studied a dependence of the response on the magnetic field at fixed radiation frequencies.  However, having in mind future experiments, in    this section  we present theoretical  calculation of the ratchet current as a function of the
 radiation frequency. The study of such dependence is very instructive, because  the Dingle factor describing rapid SdH oscillations as well as large factor $(2\pi E_F/\hbar \omega_c)^2 $ are independent of the radiation frequency. Hence, the frequency dependence clearly shows resonances  and is more convenient for  comparison of  HD and DD approaches.
We also notice that  since $\omega$  does not  enter   $\gamma'',$
the ratchet current spectrum is independent of neither $T$ nor $E_{\rm F}$.
It  is convenient to plot  the current as a function of the dimensionless  parameter $\omega/\omega_c$  and measure current in the frequency-independent units: $\tilde j_0 \delta_{\rm c}  (2\pi E_F/\hbar \omega_c)^2$  and  $ j_0 \delta_{\rm c}  (2\pi E_F/\hbar \omega_c)^2$  for HD and DD approximations, respectively.  
In order to study  in more detail the  peak at  $\omega=2 \omega_{\rm c},$ we also assume that  the second harmonic relaxes  with  a   different rate, $\tau_2 \neq \tau_{\rm tr}$, and investigate evolution  of the dc photocurrent  with variation of the ratio  $\tau_2 / \tau_{\rm tr}$. Corresponding plots are shown in Fig.~\ref{Fig6} for HD (top panel) and DD (bottom panel) regimes, respectively.
The top panel of  Fig.~\ref{Fig6}
 clearly shows CR and MP resonances at $\omega_q$ close to $\omega_c$. The splitting between the resonances increases with increasing of  the ratio $\omega_q/\omega_{\rm c}.$
 The bottom panel of  Fig.~\ref{Fig6}
  shows that the second harmonics dominates for $\tau_2$ close to $\tau_{\rm tr}.$ Importantly, the peak at the second harmonic has an asymmetric shape in contrast to conventional CR peak. With decreasing $\tau_2$ this peak disappears.

\section{Discussion and comparison of the theory with the experiment}
\label{discussion}

%
%

\begin{figure*}[t]
	\centering
	\includegraphics[width=0.9\textwidth]{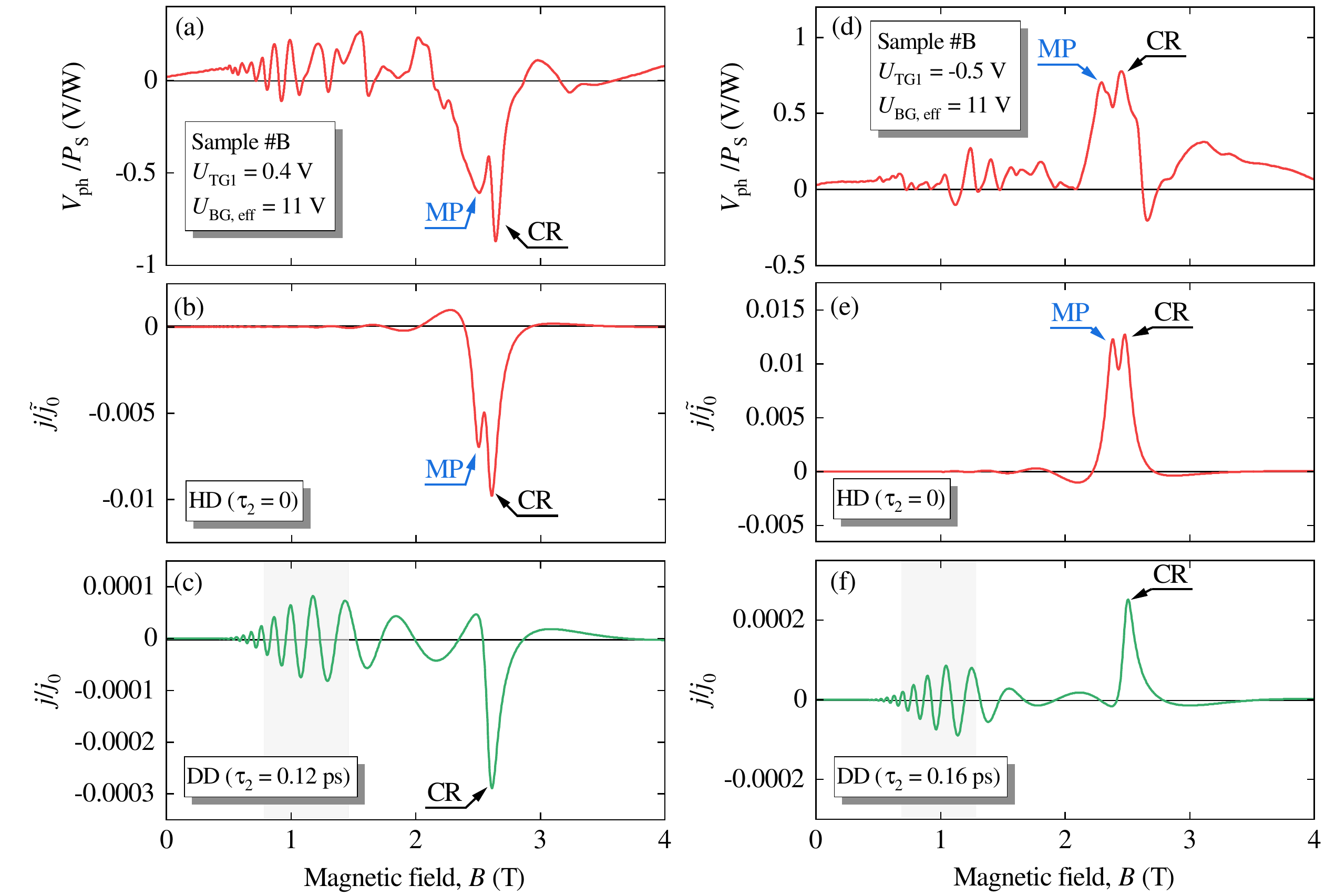}
	\caption{
 	Panels~(a), (d): Experimental results for different signs of top gate voltages show inversion of sign of the current under inversion of parameter $\Xi$.
	 Panel~(b): Theoretical curve within the HD approach ($\tau_2=0$) calculated using $\tau_{\rm tr}= 2.9$~ps, $\tau_{\rm q}=0.1$~ps, $m_{\rm eff} = 0.028 \: m_0$, 
	 $N_0=0.64 \times 10^{12}$~cm$^{-2}$,
			and $s=2.55\times10^8$~cm/s as the best parameters to fit the photoresponse of sample \#B 			 shown in panel (a). 
			Panel~(c) shows a fit within the DD approximation using the same parameters as for the HD approach and $\tau_2=0.12$~ps.
			Panel~(e): Theoretical curve within the HD approach ($\tau_2=0$) calculated using 
			the same parameters as in panel~(b) for $\tau_q$, $m$ and $s$  and $\tau_{\rm tr} = 2.6~ps$, $N_0=0.61 \times 10^{12}$~cm$^{-2}$ as the best parameters to fit the photoresponse of sample \#B shown in panel (d). Panel~(f) shows a fit within the DD approximation using the same parameters as for the HD approach and 
			$\tau_2=0.16$~ps. 		The blue and black labels indicate the position of the MP resonance and CR, respectively. The gray shaded area highlights the region for the second harmonic.
			}
	\label{Fig7}
\end{figure*}

\begin{figure}
	\centering
	\includegraphics[width=0.45\textwidth]{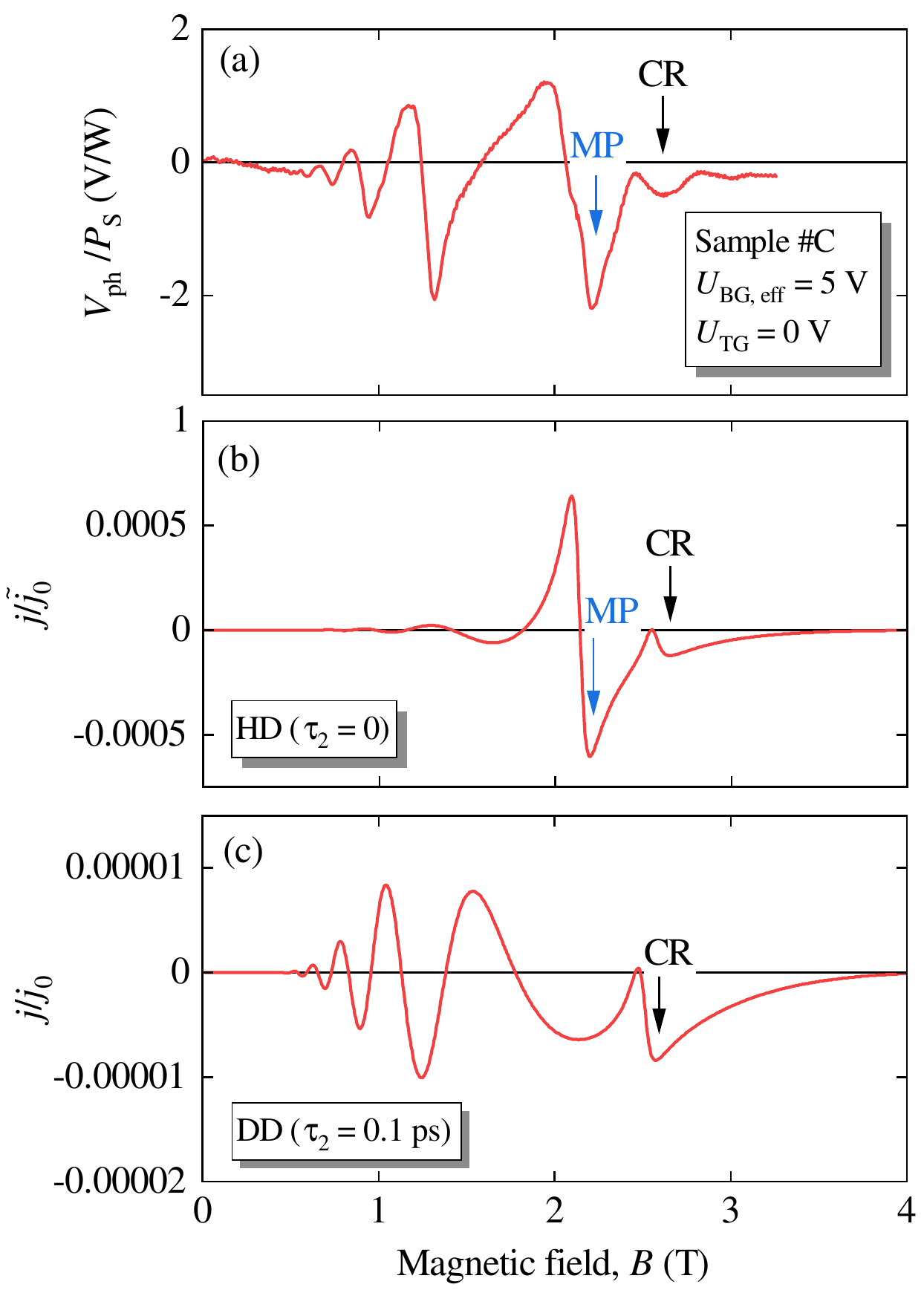}
	\caption{Theoretical results calculated with the parameters 
	$\tau_{\rm tr}= 3.1$~ps,  $\tau_{\rm q}=0.12$~ps, $m_{\rm eff} = 0.028 \: m_0$, $N_0=0.31 \times 10^{12}$~cm$^{-2}$, yielding the best fit of the photoresponse of sample \#C shown in panel~(a).  Calculations are performed within the HD approach at $s = 2.8 \times 10^8$~cm/s (panel b) and DD approach with $\tau_2=0.1$~ps (panel c). The blue and black vertical arrows show the position of the MP resonance and CR, respectively.
	}
	\label{Fig8}
\end{figure}

\begin{figure}
	\centering
	\includegraphics[width=0.45\textwidth]{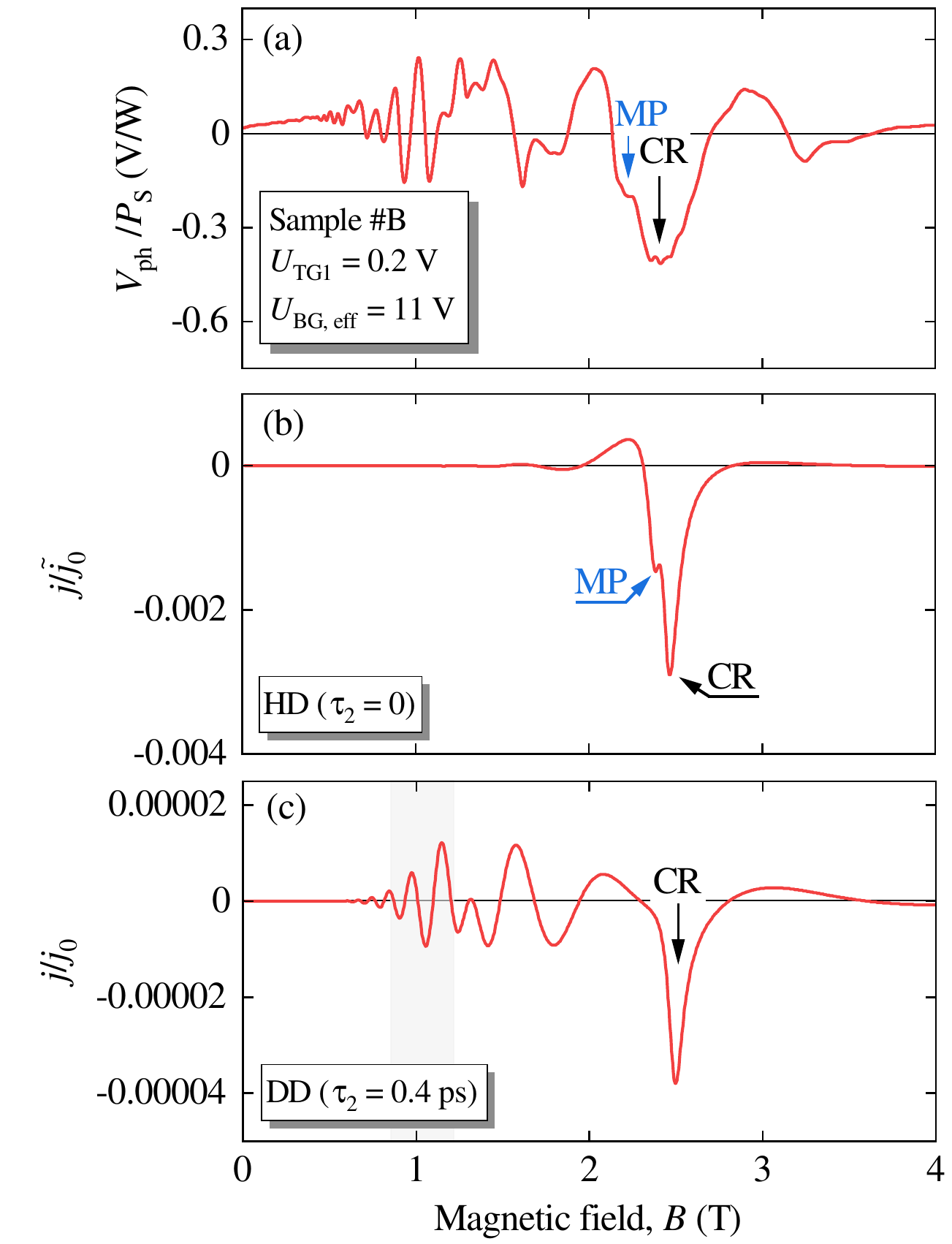}
	\caption{Theoretical results calculated with the parameters 
		$\tau_{\rm tr}= 3.1$~ps, $\tau_{\rm q}=0.065$~ps, $m_{\rm eff} = 0.027 \: m_0$, $N_0=0.62 \times 10^{12}$~cm$^{-2}$ yielding the best fit of the photoresponse of sample \#B shown in panel~(a).  Calculations are performed within the HD approach at 
		$s = 2.4 \times 10^8$~cm/s (panel b) and DD approach with 
		$\tau_2=0.4$~ps (panel c).}
	\label{Fig9}
\end{figure}

\begin{figure*}
	\centering
	\includegraphics[width=0.9\textwidth]{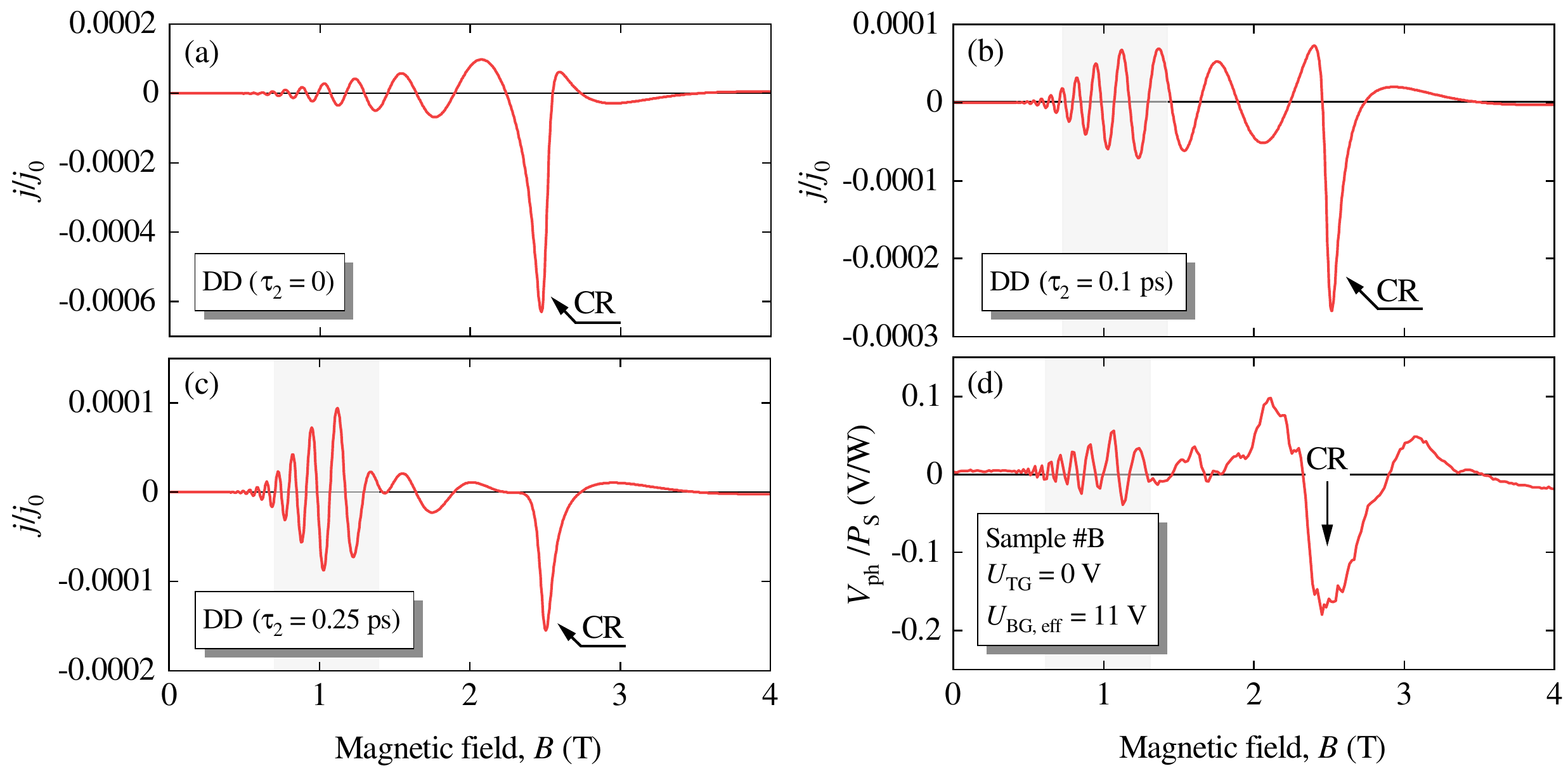}
	\caption{Panels (a) -- (c): Theoretical curves calculated within the DD approach using $\tau_{\rm tr}= 3$ ps, $\tau_{\rm q}=0.1$ ps,  $m=0.027 \: m_0,$ and $N_0=0.6 \times 10^{12}$~cm$^{-2}$	for different values of $\tau_2$. Panel (d) shows the photovoltage obtained in sample \#B. The black vertical arrows and labels indicate the position of the cyclotron resonance. The gray shaded area highlights the region for the second harmonic.  
	}
	\label{Fig10}
\end{figure*}

Now we discuss the experimental results in the view of the developed theory. Figures~\ref{Fig7}, \ref{Fig8}, and \ref{Fig9} show experimental traces previously presented in Sec.~\ref{results} together with theoretical curves calculated within both the HD and the DD approximations.
We present plots for experimental values ${T= 4.2}$~K,   $E_{\rm F}$ ranging from 25 to 100~meV and corresponding concentrations $N_0$ from 0.3 to 1.2$\times 10^{12}$ cm$^{-2}$, respectively.
First of all,  we emphasize that all traces clearly demonstrate that the polarity of the photoresponse changes upon inversion of sign of the lateral asymmetry parameter $\Xi$ [see Eq.~\eqref{Xi}], being a fingerprint of the ratchet effect. Experimentally, we change the sign of $\Xi$ by  inverting the sign of one of the top gate voltages, which leads to an inversion of the modulating static potential $V.$  
The  sign change of the current  is  illustrated in   Figs.~\ref{Fig4}(a) and \ref{Fig4}(b) and Figs.~\ref{Fig7}(a) and \ref{Fig7}(d)  and discussed in more detail in Sec.~\ref{results} and  Appendix \ref{top_gate_dep}.

The experimental data for different structures and back (top) gate voltages are shown in Figs.~\ref{Fig7}(a), \ref{Fig7}(d), \ref{Fig8}(a) and \ref{Fig9}(a). Figures~\ref{Fig7}(b), \ref{Fig7}(e),
\ref{Fig8}(b) and \ref{Fig9}(b)
depict the curves calculated in the HD regime [using Eq.~\eqref{supp-J}] 
and Figs.~\ref{Fig7}(c), \ref{Fig7}(f), \ref{Fig8}(c) and \ref{Fig9}(c) show theoretical curves for the DD regime [calculated by using Eq.~\eqref{j_tot_fin}]. For comparison of the theory and experiment we extracted several parameters used in our calculations from magnetotransport measurements, namely, the transport relaxation time $\tau_{\rm tr}$, quantum relaxation time $\tau_{\rm q}$, and the electron concentration\EMs{,} $N_0$. An important parameter of the theory is the plasma wave frequency, $\omega_q.$ A rigorous calculation of this frequency is quite  difficult because the system is inhomogeneous and contains regions with different gate voltages within one superlattice period. On the other hand, the theory was developed under the assumption that  the inhomogeneities are sufficiently small. Therefore, in order to be consistent with the theoretical assumptions, we  chose an equation for the gated homogeneous case, $\omega_q=s q,$ and used the plasma wave velocity $s$ as a fitting parameter. For all structures,  the best fit  was obtained for   $s$ being  of the order of $10^8~$cm/s. This value is very close to the experimental data on the plasma wave velocity studied in BLG  \cite{Ryzhii2020} for similar concentrations {($\sim 10^{12}$ cm$^{-2}$)} as used in our experiment. Note that although the CR position, which is defined by the effective mass and depends slightly on the back gate voltage, was also used for fine fitting, its  fit values agree well with that of the CR in unstructured BLG with similar {transport properties}, see Ref.~\cite{Candussio2021} and references therein.
Finally, the time $\tau_2$ characterizing the ee-collision rate, was used for the analysis of the resonance at the second harmonic of the CR by affecting its amplitude within the DD approximation.
 
The theoretical expressions for the photocurrent obtained in Sec.~\ref{model} are proportional to $\delta_c$,
i.e., {they} exhibit $1/B$-periodic oscillations following the SdHO, see Eqs.~\eqref{j_tot} and~\eqref{j_tot_HD}, for the DD and HD regimes, respectively.
This prediction perfectly agrees with the experiment as we demonstrated in Sec.~\ref{results} and Appendix~\ref{SdHO_analysis}, by analyzing the extrema positions of the experimentally observed photocurrent.  Furthermore, the observed giant enhancement of the photocurrent, as compared to that excited at zero magnetic field, is also in excellent agreement  with the theory. The enhancement is defined by the factor $(2\pi E_{\rm F}/\hbar\omega_c)^2$, see Eq.~\eqref{Eq15}, which, depending on the carrier density, varies in our experiments from 200 to 3000 for $B = 2$~T.

Our central experimental result is the observation of two peaks, namely the CR and MP resonances. As clearly seen from  Figs.~\ref{Fig7}, \ref{Fig8}, and \ref{Fig9}
both observed resonances are perfectly described by the HD theory. 

One of the interesting prediction of the theory is the asymmetry of the resonances. Physics behind this asymmetry is interference of homogeneous and inhomogeneous ac responses resulting in   non-linear   conversion into dc current.  The asymmetry can be easily seen from the simplified Eq.~\eqref{resonance}, which shows that split resonances are approximately symmetric for a small plasmonic splitting (i.e. for small $\Delta$), and become more and more asymmetric by an increase of  $\Delta$, {see also Fig.~\ref{Fig6}(a)}.   This is exactly what we see in the experiment. Indeed, the resonances in  Figs.~\ref{Fig7}(a) and \ref{Fig7}(d) have a much more symmetric shape than the peaks in Fig.~\ref{Fig8}(a), which correspond to a larger splitting.  Moreover, our theory also explains why the splitting is larger for Fig.~\ref{Fig8}(a). According to Eq.~\eqref{Delta}, $\Delta \propto s^2 q^2 \propto N_0 /L^2 $ (here, we take into account that $s \propto \sqrt{ N_0}$).  Figure~\ref{Fig8}(a) presents the data for sample  \#C where $L$ and $N_0$ are twice as small as compared to sample  \#B,  which is depicted in Figs.~\ref{Fig7}(a) and \ref{Fig7}(d) [$L=4~\mu $m and $L=2~\mu$m for  \#B and  \#C, respectively]. Hence, the splitting in sample  \#C should be approximately twice larger than in sample  \#B  as demonstrated in the experiment.

Now we comment on the applicability of the DD approach to the description of the experimental data. Figures~\ref{Fig7}(c),  \ref{Fig7}(e),  \ref{Fig8}(c), and \ref{Fig9}(c) demonstrate that while the  presented theory of the DD regime describes the CR well, the  MP resonance is missing. The MP resonance however can also be obtained in the DD approach if one includes long-range ee-interaction in the DD approach, which is a task for the future.

As an important result, the DD approach does not only describe the CR resonance but also explores the observed amplification of the ratchet signal at the second harmonic of the CR, which is absent in the HD approach. Specifically, the  resonance at $ \omega_c=\omega/2$ is not captured by the HD approximation for an ideal electronic fluid, where  $\tau_2=0,$ and is described for our samples theoretically within the DD theory by using $\tau_2$ as a fitting parameter.
Such approach perfectly explains the observed enhancement of the oscillating magneto-ratchet response  near half the value of the $B_{\rm CR}$ field, see  experimental data and theoretical calculations considering the DD approach depicted in Figs.~\ref{Fig7}, \ref{Fig8} and \ref{Fig9}.  

In order to explore the influence of $\tau_2$  in more detail, we plotted the results of the DD calculations for different values of $\tau_2$ in Fig.~\ref{Fig10}.  From this comparison it is clearly seen that to obtain a good agreement it requires to use values for $\tau_2$, which are at least one order of magnitude shorter than the momentum relaxation time $\tau_{\rm tr}.$  The best fits for all plots in Figs.~\ref{Fig7}, \ref{Fig8},  and \ref{Fig9} are obtained using $\tau_2 \approx \tau_q \approx \tau_{\rm tr}/30$.
This implies that the ee-collisions dominate over impurity scattering, and  our system is indeed quite close to an ideal fluid.
Therefore, the HD approach is more applicable for the description of the experimental data. However, since the second harmonic although being small is clearly observed,   the electronic fluid is not fully ideal. In particular, it has a finite viscosity $\eta$, which (for zero $B$) can be estimated according to  $\eta=v_F^2 \tau_2/4$ using  experimental  values.   Taking $E_{\rm F}=100$~meV and $\tau_2=0.1$~ps, we obtain $\eta \approx 0.03$~m$^2$/s  which is  smaller but 
has the same order of magnitude as
the zero-field value of the viscosity $\eta \approx 0.1$ m$^2$/s reported in Ref.~\cite{Berdyugin2019}. We note, however, that such a value of the viscosity is obtained for  liquid He temperature implying that the ee-scattering is more significant in  our high frequency  and high magnetic  field experiments.   We also 
notice that for for experimental values $\tau_{\rm tr} \approx 3$ ps  and $L=4~\mu$m,  viscous damping $\eta q^2 $ is smaller as compared  to $1/\tau_{\rm tr}$ as was initially assumed in Eq.~\eqref{v}.
  The HD nature of the electron transport in our samples at 
$T=4.2$~K is strongly supported by the previous analysis of the frequency dependence of the ratchet current at $B=0$~\cite{Moench2022}.
Note that the second harmonic could be even more pronounced (as compared to the first one) at sufficiently higher temperatures {~\cite{Moench2022}}, when electron-phonon interactions 
{drive} the system into the DD regime. However, at such temperatures that SdH oscillations are suppressed.

Before closing the discussion, we briefly comment on other features of the ratchet current. First of all, we notice strong frequency dependence of the effect. This dependence  is clearly seen from the theoretical  curves shown in  Fig.~\ref{Fig6}. 
Detailed experimental investigation  of the frequency dependence is a challenging {task involving} a great number of additional experiments and {requires further study}.  The key obstacle  here is the absence   of  a tunable THz source in a wide range of frequencies. Moreover, for lower frequencies the CR and  MP resonances have low quality factors and are {concealed by} the dense SdHO. However, in Appendix~\ref{frequencies} we present some  preliminary experimental  results  showing  the evolution  of  the    photocurrent dependence on the magnetic field with the frequency decrease.

We do not focus in this work on the polarization  dependence  of the signal. 
{However, we notice}  that  the theoretically   predicted dependence on polarization  qualitatively agrees with the obtained data. 
For linearly polarized radiation, the  ratchet current consists of the polarization-independent current as well as of two contributions varying upon rotation of the radiation polarization plane as $P_{\rm L1}=\cos{2\alpha}$ and $P_{\rm L2}= \sin{2\alpha}$,  see Eqs.~\eqref{j_tot} for the DD regime and~\eqref{j_tot_HD} for the HD regime. As addressed in Sec.~\ref{results}, these contributions are clearly detected in the experiment demonstrating that the total photosignal is described by $V_{\rm ph} = V_0 + V_{\rm L}\cos(2\alpha + \theta)$.

Experiments also clearly show a contribution of the magnetic ratchet current $V_{\rm C} \propto P_{\text{C}}$, which changes its sign upon switching the radiation helicity, see Fig.~\ref{Fig5}. The helicity-driven current is also expected from the developed theory, see last terms on the right sides of Eqs.~\eqref{j_tot} and~\eqref{j_tot_HD}. These formulas also show CR and MP  resonances.  A signature of these resonances is shown in  Fig.~\ref{Fig5} {labeled} by gray arrows. More detailed study of the frequency and polarization of the response will be presented elsewhere.

\section{ Summary}
\label{summary}

To summarize, we explored, both theoretically and experimentally, {\it resonant} ratchet effects arising in the magnetic field and investigated the role of the ee-interaction. Specifically, we observed a tunable --- by   magnetic field  and gate voltage --- {\it resonant} conversion  of the terahertz radiation into a dc current in a BLG sample superimposed with a lateral superlattice formed by a DGG structure. The resonant dc photoresponse was observed in the SdH regime, where the ratchet current exhibits giant sign-alternating
magneto-oscillations.

Our key findings are the direct observation of a sharp CR in the photocurrent, which dramatically enhances the envelope of SdH oscillations, and also the demonstration of two effects caused by ee-interaction:
\begin{enumerate}
\item[(i)] a plasmonic splitting of the CR onto two peaks due to long-range Coulomb interaction;
\item[(ii)] a strong suppression of the second harmonic of the CR due to fast ee-collisions.
\end{enumerate}

We developed  a  theory which  perfectly fits our data and describes the  resonance splitting as a  co-existence of the CR at $\omega=\omega_c$  and a MP satellite Eq.~\eqref{MP}. The theory shows that the possibility of such a co-existence, in contrast to  naively expected  simple plasmonic shift of the CR,  is a remarkable feature of the non-linear ratchet effect. Although we present detailed comparison with experiment only for linear polarization perpendicular to the grating, the theory qualitatively  reproduces experimentally observed polarization  dependence.   

Moreover, the theory  predicts  the resonance  at the  second harmonic of CR, $\omega=2 \omega_c,$  which was clearly seen in the experiment, but with a strongly suppressed amplitude.  We associate this suppression to fast ee-collisions and conclude that our system is in  the HD regime at  liquid He temperature. This conclusion is in an excellent agreement with our recent study of the same structures at zero magnetic field.

\section*{Acknowledgments}

The support of  the Deutsche Forschungsgemeinschaft (DFG, German Research Foundation) project Ga501/18, RFBR project 21-52-12015, IRAP  Programme  of the Foundation   for   Polish Science   (grant   MAB/2018/9, project CENTERA), and the Volkswagen Stiftung Program (97738) is gratefully acknowledged. J.E. acknowledge support of the DFG through SFB 1277 (project-id 314695032, subproject A09) and GRK 1570. The work of V.Yu.K. and S.O.P. was  also supported  by RFBR grant No. 20-52-12019.
L.E.G.  and S.O.P. also thank the Foundation for the Advancement of Theoretical Physics and Mathematics ``BASIS''. Growth of hexagonal boron nitride crystals was supported by JSPS KAKENHI (Grant Numbers 19H05790, 20H00354 and 21H05233).

\appendix \section{Experiment}
\label{Supplemental_exp}

\subsection{Ratchet-currents at different frequencies}
\label{frequencies}

\begin{figure}
	\centering
	\includegraphics[width=\linewidth]{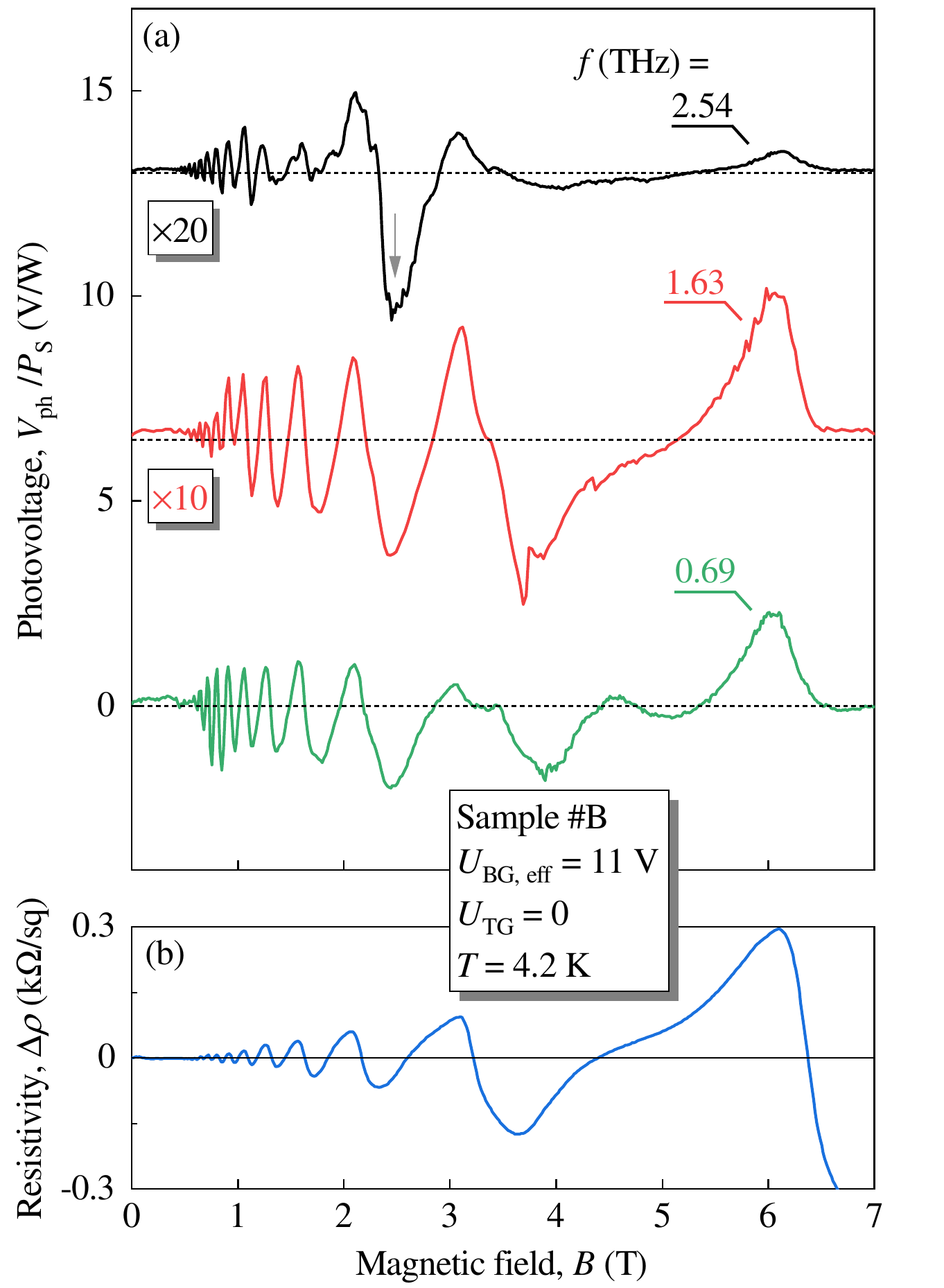}
	\caption{Panel (a): Magnetic filed dependence of the magneto-ratchet in sample \#B obtained for three different THz frequencies, $f = 2.54$, 1.63 and 0.69~THz. All traces were measured for high carrier density, $U_{\rm BG, eff} = 11$~V, whereas the top gates were kept at zero bias, $U_{\rm TG} = 0$. The polarization state was directed perpendicular to the metal stripes of the DGG ($\alpha = 0$) throughout all frequencies.
	The curves for 2.54 and 1.63~THz were multiplied by factors as labeled and are offset for clarity by 6.5 V/W. The gray  vertical arrow depicts the CR position for $f = 2.54$~THz, see Fig.~\ref{Fig10}. Panel~(b): Magnetotransport dependence of the oscillating part of sample's resistivity $\Delta\rho$ measured under the same conditions. 
		}
	\label{Fig11}
\end{figure}

Figure~\ref{Fig11}(a) shows magnetic field dependencies of the ratchet effect obtained in sample \#B for a high carrier density and three different frequencies $f = 2.54$, 1.63 and 0.69~THz. All three traces clearly show 1/$B$-periodic magnetooscillations with an amplitude substantially larger as the signal at zero field. Comparison of these traces with the oscillatory part of the magnetoresistivity $\Delta\rho$, plotted in panel (b), demonstrates that the oscillation period is equal to that of the SdHO. While the oscillation behavior is the same for all three frequencies the amplitude of the signal strongly increases with the frequency decrease, see factor for different traces in Fig.~\ref{Fig11}(a). One more difference is the appearance of the CR and MP resonances as well as amplitude enhancement at $B = B_{\rm CR}/2$ for $f = 2.54$~THz. For traces at lower frequencies we do not see any specific features at this magnetic field strength. As discussed in the main text, the position of the magnetic field addressed above depends linearly on the radiation frequency, thus, for lower frequencies they are expected for substantially lower magnetic fields. The analysis of the resonances at such fields is complicated by the dense SdH-like oscillations 
and lower quality factors of the resonances.

\subsection{SdH-like oscillations in photovoltage}
\label{SdHO_analysis}

\begin{figure*}
	\centering
	\includegraphics[width=\linewidth]{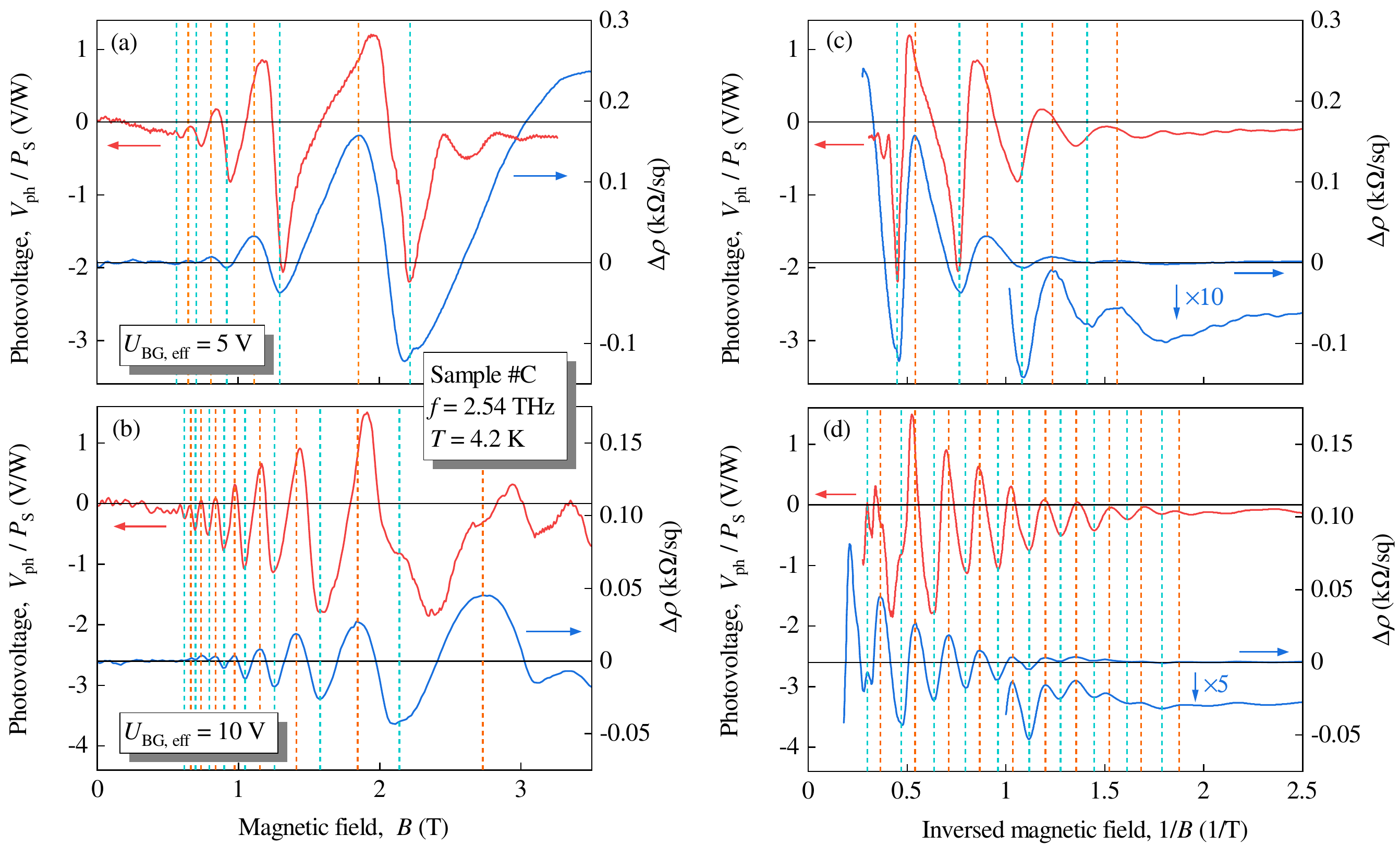}
	\caption{Photovoltage normalized by the radiation power as a function of magnetic field (panels a and c) and inverse magnetic field (panels b and d) obtained in sample \#C for two effective back gate voltages, $U_{\rm BG, eff} = 5$ and 10~V, keeping both top gates at zero bias ($U_{\rm TG} = U_{\rm TG1} = U_{\rm TG2} = 0$). The data are obtained for $f = 2.54$~THz with linear polarization ($\alpha = \Degree{90}$). Blue lines show corresponding dependencies of the oscillating part of sample's resistivity $\Delta\rho$ measured under the same conditions. Low signal parts in the left panels are zoomed by multiplication with factors 10 and 5 in panels (b) and (d), respectively. Vertical dashed lines indicate extreme positions of the SdH oscillations in the resistivity traces. The black arrow labeling CR indicates the cyclotron resonance position in the photosignal. The photovoltage traces were multiplied by -1 for better comparison with the resistivity data. 
	}
	\label{Fig12}
\end{figure*}

Red traces in Fig.~\ref{Fig12} show the $B$-field and 1/$B$-field dependencies of ratchet signal measured in sample \#C at two back gate voltages $U_\mathrm{BG, eff} = 5$ and 10~V. They demonstrate that the ratchet signal exhibits sign-alternating magneto-oscillations with an amplitude by more than an order of magnitude larger than the photosignal at zero magnetic field. Comparing these oscillations with the oscillating part of the magnetoresistivity we obtained that, the extreme positions of the oscillations of ratchet effect coincide with that of the SdHO (blue curves in Fig.~\ref{Fig12}). This is indicated by the vertical dashed lines (orange for maximum and cyan for the minimum) in Fig.~\ref{Fig12}.

\subsection{Top gate dependence}
\label{top_gate_dep}

Here, we discuss data on top gate dependence which prove that observed dc current  is caused by ratchet effect. As we already mentioned in the main text, Figs.~\ref{Fig7}(a) and (d) compare results for opposite signs of the lateral asymmetry parameter $\Xi$ given by Eqs.~\eqref{Xi0} and~\eqref{Xi}. As addressed in the main text, the sign reversal of the magneto-photocurrent is in a full agreement with the theory yielding that the ratchet photocurrent is proportional to $\Xi$, see Eqs.~\eqref{j_tot}  and~\eqref{j_tot_HD} where $j_0 \re{, \tilde j_0}\propto \Xi$, Eq.~\eqref{j0}. Note that besides the data for two fixed sequences of the voltages applied to the top gates TG1 and TG2, we also varied one of the top gate voltages keeping the other at zero bias. Note that for both top gate voltages equal to zero, the magnetic ratchet currents are caused by the built-in asymmetry.
 
Below we present  some additional data 
showing that the sign and amplitude of the ratchet photocurrent can  be easily controlled   by  the top gate potentials.

\begin{figure}
	\centering
	\includegraphics[width=0.85\linewidth]{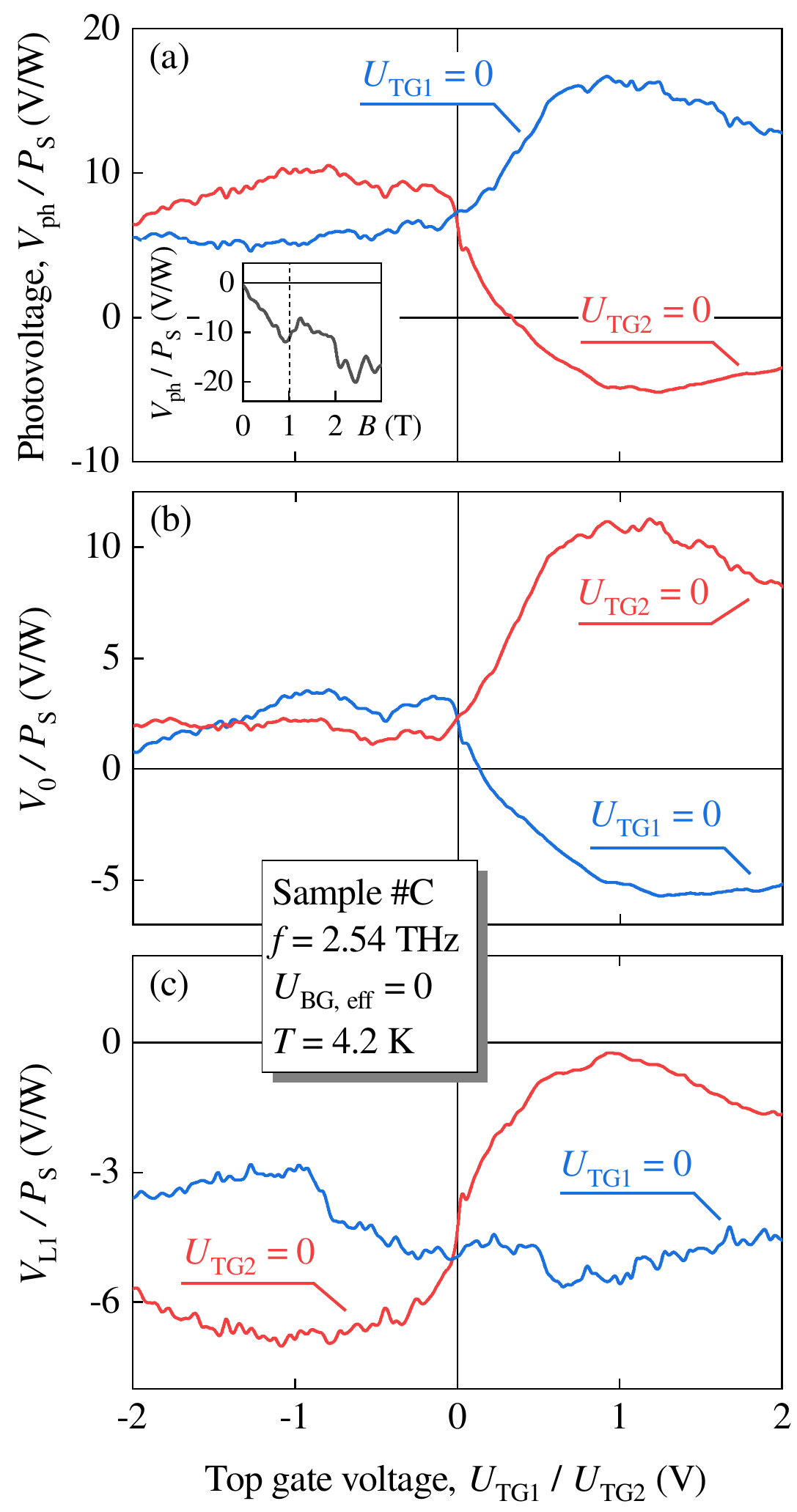}
	\caption{Panel (a) illustrates the top gate voltage dependencies of the magneto-ratchet current measured in sample \#C at a fixed magnetic field $|B| = 1$~T and $U_{\rm BG, eff} =0$. The red (blue) trace corresponds to a TG1 (TG2) sweep and was recorded while the other top gate was held at zero bias. The inset shows the photovoltage measured for low carrier density as a function of the magnetic field for $U_{\rm TG1}/U_{\rm TG2} = 2$~V/0. Top gate voltage dependencies of (b) linear $V_{\rm L1}$ and (c) polarization insensitive $V_{\rm 0}$ ratchet contributions measured for sample \#C. The individual contribution has been extracted from two sets of measurements applying THz radiation of frequency $f = 2.54$~THz polarized linearly at the azimuth angles $\alpha =0$ and \SI{90}{\degree} using $V_\mathrm{L1} = \left[V_\mathrm{ph}(\alpha = \SI{0}{\degree}) - V_\mathrm{ph}(\alpha = \SI{90}{\degree})\right]/2$ and $V_\mathrm{0} = \left[V_\mathrm{ph}(\alpha = \SI{0}{\degree}) + V_\mathrm{ph}(\alpha = \SI{90}{\degree})\right]/2$. 
	}
	\label{Fig13}
\end{figure}

In particular, the top gate dependencies at rather small constant magnetic field at which the 1/$B$ oscillations are not pronounced, yet, see inset in Fig.~\ref{Fig13}(a), are studied to support additionally that the sign and the amplitude of the photoresponse are determined by the lateral asymmetry parameter $\Xi$. For that in experiments we fixed one top gate at zero voltage and varied the other one from negative to positive values. Experiments were carried out for linearly polarized radiation with $\alpha = 0$ and \SI{90}{\degree}. Figure~\ref{Fig13}(a) shows top gate dependencies of the total photovoltage excited by radiation electric field oriented perpendicularly to the DGG stripes. It is seen that TG1/TG2 dependencies are almost horizontally mirrored. Note that at zero top gate voltages the signal is rather strong, which is caused by a large built-in asymmetry due to the presence of metal stripes on top of encapsulated BLG~\cite{Moench2022, Olbrich2016}. Due to this fact, the reversing of top gate voltage polarity changes the sign of the ratchet signal for one traces only at $U_{\rm TG1}/U_{\rm TG2} > 0$. Using the data for the two azimuth angle $\alpha$ we extracted top gate dependencies for polarization independent $V_0$ and polarization sensitive $V_{\rm L1}$ contribution, see panel (b) and (c) in Fig.~\ref{Fig13}, respectively. 

 \section{Theory}
\label{Supplemental}

\subsection{Derivation of the ratchet current in the DD regime}

The ratchet current Eq.~\eqref{j_tot} is obtained by sequential iterations of the kinetic equation~\eqref{kin_eq} at two small perturbations, namely the light amplitude $\bm E$ and the periodic ratchet potential $V(x)$. The first iteration step is always account for $\bm E$, but the next steps could be different. One contribution is obtained if the potential $V$ is taken into account at the second stage, and the radiation amplitude $\bm E$ at the last stage. We denote the corresponding correction to the distribution function $f^{(EVE)}$. 
Similar to systems with linear energy dispersion~\cite{Hubmann2020}, the total ratchet current is not restricted to this contribution.
An additional contribution to the ratchet current, $\delta \bm j$, is obtained if the amplitude $\bm E$ is taken into account twice assuming $V=0$, and then, at the last stage, the periodic potential is taken into account. The corresponding part of the distribution function is denoted as $f^{(EEV)}$. Below we derive both contributions to the ratchet current.

\subsubsection{$EVE$ contribution}
\label{App_EVE}

The distribution function $f^{(EVE)}$ is a solution of the kinetic equation bilinear in $\bm E$ and linear in $V(x)$ obtained by a simultaneous account for $\bm E$, $V(x)$, and then $\bm E$. 

The kinetic equation for $f^{(EVE)}$ has the form
\begin{equation}
\omega_c {\partial f^{(EVE)} \over \partial  \varphi} + e \bm E^* \cdot {\partial f^{(EV)} \over \partial \bm p} = 
\text{St}\qty[f^{(EVE)}],
\end{equation}
where $f^{(EV)}$ is the correction bilinear in $\bm E$ and $V(x)$,
and the collision integral $\text{St}[f]$ is characterized by the relaxation rates of the Fourier harmonics
\begin{equation}
\gamma_i(\varepsilon)={1\over\tau_i} \left[ 1+\delta_c(\varepsilon) \right], \qquad i=1,2.
\end{equation}
At low temperatures $\tau_1$ coincides with the transport relaxation time $\tau_{\rm tr}$ which determines the mobility.

Solution is given by
\begin{equation}
\label{f_EVE}
f^{(EVE)} = - \sum_\pm \tau_c^\pm e \bm E^* \cdot \left( {\partial f^{(EV)} \over \partial \bm p} \right)_\pm + c.c.,
\end{equation}
where $(\ldots)_\pm$ denotes the $\pm 1$st Fourier-harmonics, and 
\begin{equation}
\tau_c^\pm = {1 \over \gamma_1(\varepsilon) \pm i\omega_c}.
\end{equation}
Here $\delta_c(\varepsilon)$ means $\delta_c$ given by Eq.~\eqref{delta_c} where $E_\text{F}$ is substituted by $\varepsilon$.

Substituting the solution~\eqref{f_EVE} into Eq.~\eqref{j}, we obtain the current density in the form
\begin{equation}
j_\alpha = -e^2\sum_{\nu,\bm p} v_\alpha \sum_\pm \tau_c^\pm  \bm E^* \cdot \left( {\partial f^{(EV)} \over \partial \bm p} \right)_\pm + c.c.
\end{equation}
This equation shows that only the even in $\bm p$ part of $f^{(EV)}$ contributes to the photocurrent. It contains two terms, the $\varphi$-independent one and the 2nd harmonics of $\varphi$.
For $j_+=j_x+ij_y$ we get
\begin{equation}
j_+ = -e^2\sum_{\nu,\bm p} v_+  \tau_c^-  \bm E^* \cdot \left( {\partial f^{(EV)} \over \partial \bm p} \right)_- \\+ (\bm E \leftrightarrow \bm E^*, \:  \omega \rightarrow -\omega).
\end{equation}
Integrating by parts we obtain
\begin{equation}
j_+ = e^2\sum_{\nu,\bm p} {\partial (v_+  \tau_c^-)\over \partial \bm p}  \cdot  \bm E^* f^{(EV)} + (\bm E \leftrightarrow \bm E^*, \: \omega \to -\omega).
\end{equation}
Calculating the gradient in the momentum space
\begin{equation}
{\partial (v_+  \tau_c^-)\over \partial \bm p}  \cdot  \bm E^* 
= {1\over m} \left[(E^*)_+ (\varepsilon \tau_c^-)' + (E^*)_-\varepsilon(\tau_c^-)'\text{e}^{2i\varphi_{\bm p}} \right], 
\end{equation}
where $(E^*)_\pm = E^*_x\pm iE^*_y$,
and the prime denotes differentiating over $\varepsilon$,
we obtain
\begin{widetext}
	\begin{equation}
	j_+ = {e^2\over m} \sum_{\nu,\bm p} \left[(E^*)_+ (\varepsilon \tau_c^-)'f^{(EV)}_0 + (E^*)_-\varepsilon(\tau_c^-)'f^{(EV)}_{-2} \right]  + (\bm E \leftrightarrow \bm E^*, \: \omega \to -\omega).
	\end{equation}
	Here $f^{(EV)}_{0,-2}$ mean the angular-independent part of $f^{(EV)}$ and the part $\propto \text{e}^{-2i\varphi_{\bm p}}$.
	
	Since the angular integration is already performed,  we can pass from summation over $\bm p$ to integration over energy:
	\begin{equation}
	\label{j+}
	j_+ = {e^2\over m} \int \dd\varepsilon g(\varepsilon)  \left[(E^*)_+ (\varepsilon \tau_c^-)'f^{(EV)}_0 + (E^*)_-\varepsilon(\tau_c^-)'f^{(EV)}_{-2} \right]  + (\bm E \leftrightarrow \bm E^*, \: \omega \to -\omega),
	\end{equation}
	where $g(\varepsilon)$ is the density of states (with account for spin and valley degeneracies).
\end{widetext}
The corrections $f^{(EV)}_{0,-2}$ are given by
\begin{equation}
f^{(EV)}_0 = 
{ie\over 2m\omega} \sum_\pm \left[-f_0' E_\mp{dV\over dx} \left(\varepsilon \tau_{1\omega}^\pm  \right)' + V{d E_\mp\over dx}\varepsilon \tau_{1\omega}^\pm f_0''\right],
\nonumber
\end{equation}
\begin{equation}
f^{(EV)}_{-2} 
= {e\varepsilon \tau_{2\omega}^-\over 2m}  \left[-f_0' E_+{dV\over dx}  \left(\tau_{1\omega}^- \right)' + V{d E_+\over dx}  \tau_{1\omega}^- f_0''\right],
\nonumber
\end{equation}
where 
\begin{equation}
\label{tau_n_omega}
\tau_{n\omega}^\pm = {1\over \gamma_n(\varepsilon)-i\omega \pm in\omega_c},
\quad n=1,2.
\end{equation}
These expressions at $\omega_c=0$ pass into the corresponding expressions from Ref.~\cite{Nalitov2012}.

Substituting $f^{(EV)}_{0,2}$ into Eq.~\eqref{j+}, we obtain 
\begin{equation}
\label{j+tot}
j_+ = j_+^{(0)} +j_+^{(-2)}+ (\bm E \leftrightarrow \bm E^*, \: \omega \to -\omega),
\end{equation}
where
\begin{widetext}
	\begin{equation}
	j_+^{(0)} = {ie^3\over 2m^2\omega}  \sum_\pm \int \dd\varepsilon  g(\varepsilon) \left(\varepsilon \tau_c^-\right)'  (E^*)_+  \left[-f_0' E_{\mp}{dV\over dx} \left(\varepsilon \tau_{1\omega}^\pm  \right)'  + V{d E_{\mp}\over dx}  \varepsilon\tau_{1\omega}^\pm f_0''\right],
	\end{equation}
	\begin{equation}
	j_+^{(-2)} ={e^3 \over 2m^2} \int \dd\varepsilon g(\varepsilon) \varepsilon^2\tau_{2\omega}^- \left(\tau_c^-\right)'   (E^*)_-   \left[-f_0' E_+{dV\over dx} \left(\tau_{1\omega}^-\right)' + V{d E_+\over dx}  \tau_{1\omega}^- f_0''\right].
	\end{equation}
	Averaging over the $x$ coordinate with $E_0$ being the near-field amplitude yields
	\begin{equation}
	\overline{E_0V{d E_0\over dx}} ={1\over 2}\overline{V{dE_0^2 \over dx}} =  -{1\over 2}\overline{E_0^2 {dV\over dx}} \equiv -{1\over 2}\Xi,
	\end{equation}
	and integrating over $\varepsilon$  we get
	\begin{equation}
	j_+^{(0)} = \Xi 
	{ie^3\over 2m^2\omega} 
	\sum_\pm [|e_x|^2+ie_xe_y^* {\pm(|e_y|^2-ie_x^*e_y)}]
	\left\{  g  \left(E_\text{F}\tau_c^- \right)' \left(E_\text{F} \tau_{1\omega}^\pm  \right)'
	-{1\over 2} \left[ g \left(E_\text{F}\tau_c^- \right)' E_\text{F} \tau_{1\omega}^\pm \right]' \right\} ,
	\end{equation}
	\begin{equation}
	j_+^{(-2)} = \Xi (1-P_\text{C}) {e^3 \over 2m^2} 
	\left\{ g  E_\text{F}^2 \tau_{2\omega}^- \left(\tau_c^- \right)'\left(\tau_{1\omega}^- \right)'
	-{1\over 2} \left[g  E_\text{F}^2 \tau_{2\omega}^-\left(\tau_c^-\right)' \tau_{1\omega}^- \right]' \right\}.
	\end{equation}
	Here the prime denotes differentiation over $E_\text{F}$.
\end{widetext}

Let us analyze the terms in the curly brackets. The maximal result comes from the second derivative $(\tau_c^-)''$, therefore, only the second terms in curly brackets are important. The terms with the first derivative have much smaller amplitude due to the factor ${\hbar\omega_c/(2\pi E_\text{F}) \ll 1}$. The terms $\sim \delta_c^2$ are also omitted because they have an additional small factor $\exp(-\pi/\omega_c\tau_q)\ll 1$ and result in oscillations with double period not present in the experiment. As a result, we obtain
\begin{equation}
j_+^{(0)} = -\Xi {e^3 E_\text{F}^2 g\over 2m^2}  (\tau_c^-)''
i{\tau_{1\omega}^+(1+P_\text{C}) + \tau_{1\omega}^-(P_{\rm L1}+i{P}_{\rm L2})\over 2\omega},
\end{equation}
\begin{equation}
j_+^{(-2)} = - \Xi (1-P_\text{C}) {e^3E_\text{F}^2g\over 4m^2}  
Q_\omega (\tau_c^-)'',
\end{equation}
where ${g=2m/(\pi\hbar^2)}$ is the zero-field density of states (spin and valley degeneracies are taken into account), and
\begin{equation}
Q_\omega = \tau_{2\omega}^- \tau_{1\omega}^-.
\end{equation}

Finally, from Eq.~\eqref{j+tot} we obtain
the total current
\begin{align}
\label{j_fin}
j_+  & = -\Xi{e^3 E_\text{F}^2 \over \pi \hbar^2 m} (\tau_c^-)''\\
&\times \left[Q_+ - Q_-P_\text{C} + iT_0 + iT_\text{C}P_\text{C} + iT_{\rm L}(P_{\rm L1} +i{P}_{\rm L2}) \right] 
.\nonumber
\end{align}
Here 
$Q_\pm=(Q_\omega \pm Q_{-\omega})/2$
and
\begin{equation}
T_0 = {\tau_{1\omega}^+ \over 2\omega} + (\omega \to -\omega)= {\tau_{1\omega}^+ - \tau_{1,-\omega}^+\over 2\omega} ,
\end{equation}
\begin{equation}
T_\text{C} = {\tau_{1\omega}^+ \over 2\omega} - (\omega \to -\omega)= {\tau_{1\omega}^+ + \tau_{1,-\omega}^+\over 2\omega},
\end{equation}
\begin{equation}
T_\text{L} = {\tau_{1\omega}^- \over 2\omega} + (\omega \to -\omega)={\tau_{1\omega}^- - \tau_{1,-\omega}^-\over 2\omega},
\end{equation}
At circular polarization $P_\text{C}=\pm 1$ we have $j_+ \propto Q_{\mp\omega}, (T_0\pm T_\text{C})\propto \tau^-_{\mp\omega}, \tau^+_{\pm\omega}$ with CR at $\omega_c><0$.

\subsubsection{$EEV$ contribution}
\label{App_EEV}

Here we calculate the  correction $\delta \bm j$ obtained by twice account for $\bm E$ and then for $V(x)$.
The corresponding ratchet current is given by
\begin{equation}
\label{j_def}
\delta j_\alpha = e\sum_{\nu,\bm p} v_\alpha f^{(EEV)}.
\end{equation}

The kinetic equation for $f^{(EEV)}$ has the form
\begin{equation}
\omega_c {\partial f^{(EEV)} \over \partial  \varphi} - {dV\over dx} {\partial f^{(EE)} \over \partial p_x} = - \gamma f^{(EEV)},
\end{equation}
where $f^{(EE)}$ is the correction bilinear in $\bm E$.
Solution is given by
\begin{equation}
\label{f_EEV}
f^{(EEV)} = \sum_\pm \tau_c^\pm {dV\over dx} \left( {\partial f^{(EE)} \over \partial p_x} \right)_\pm.
\end{equation}

Substituting the solution~\eqref{f_EEV} into Eq.~\eqref{j_def}, we obtain the current density in the form
\begin{equation}
\delta j_\alpha = e{dV\over dx} \sum_{\nu,\bm p} v_\alpha \sum_\pm \tau_c^\pm \left( {\partial f^{(EE)} \over \partial p_x} \right)_\pm.
\end{equation}

For $\delta j_+=\delta j_x+i\delta j_y$ we get
integrating by parts
\begin{equation}
\delta j_+ = -e{dV\over dx} \sum_{\nu, \bm p} {\partial (v_+  \tau_c^-)\over \partial p_x}  f^{(EE)}.
\end{equation}

Calculating the derivative
\begin{equation}
{\partial (v_+  \tau_c^-)\over \partial p_x}   
= {1\over m} \qty[ \qty(\varepsilon\tau_c^-)'+  \varepsilon\left(\tau_c^-\right)'\text{e}^{2i\varphi_{\bm p}} ]  ,
\end{equation}
we obtain
\begin{equation}
\delta j_+ = -{e \over m}{dV\over dx} \sum_{\nu, \bm p} \qty[\qty(\varepsilon\tau_c^-)'f^{(EE)}_0 + \varepsilon\left(\tau_c^-\right)'f^{(EE)}_{-2} ] .
\end{equation}
Here $f^{(EE)}_{0,-2}$ mean the angular-independent part of $f^{(EE)}$ and the part $\propto \text{e}^{-2i\varphi_{\bm p}}$. The former is controlled by energy relaxation processes: $f^{(EE)}_0\propto \tau_\varepsilon$ with $\tau_\varepsilon$ being the energy relaxation time. It describes the Seebeck and Nernst-Ettingshausen ratchet effects~\cite{Budkin2016a}. In what follows we omit this contribution concentrating on polarization-dependent ratchet currents.

Since the angular integration is already performed,  we can pass from summation over $\bm p$ to integration over energy:
\begin{equation}
\label{d_j+}
\delta j_+ = -{e\over m}{dV\over dx} \int \dd\varepsilon g(\varepsilon) \varepsilon\left(\tau_c^-\right)' f^{(EE)}_{-2}.
\vspace{2mm}
\end{equation}
The correction  $f^{(EE)}_{-2}$ is multiplied by $dV/dx$, therefore we find it in the quasi-homogeneous limit:
\begin{equation}
-2i\omega_c f^{(EE)}_{-2} + e\bm E^*\cdot \left( {\partial f^{(E)} \over \partial \bm p} \right)_{-2} = - \gamma_2 f^{(EE)}_{-2}, 
\end{equation}
where the linear in $\bm E$ correction to the distribution function is found from
\begin{equation}
\omega_c {\partial f^{(E)} \over \partial  \varphi}+ e\bm E\cdot\bm v f_0' = - {f^{(E)}\over \tau_{1\omega}}.
\end{equation}
The solutions are:
\begin{equation}
f^{(E)} 
= -{e\over 2}\sum_\pm f_0' \tau_{1\omega}^\pm v_\pm E_\mp,
\end{equation}
\begin{equation}
f^{(EE)}_{-2} = - \tau_{c2} e\bm E^*\cdot \left( {\partial f^{(E)} \over \partial \bm p}\right)_{-2} + (\bm E \leftrightarrow \bm E^*, \: \omega \to -\omega),
\end{equation}
where $\tau_{1\omega}^\pm$ is given by Eq.~\eqref{tau_n_omega}, and
\begin{equation}
\tau_{c2}  = {1\over \gamma_2(\varepsilon)-2i\omega_c}.
\end{equation}
Calculation is performed as follows:
\begin{equation}
\sum_\pm \bm E^*\cdot \left( {\partial   \over \partial \bm p}f_0' \tau_{1\omega}^\pm v_\pm E_\mp \right)_{-2}
= |E|^2(P_{\rm L1}+i{P}_{\rm L2}){\varepsilon\over m}  \left( \tau_{1\omega}^- f_0' \right)' ,
\end{equation}
which yields
\begin{equation}
f^{(EE)}_{-2} = {e^2\tau_{c2}\varepsilon\over 2m}  |E|^2(P_{\rm L1}+i{P}_{\rm L2}) \left[ (\tau_{1\omega}^- + \tau_{1,-\omega}^-)f_0' \right]'.
\end{equation}

Substituting $f^{(EE)}_{-2}$ into Eq.~\eqref{d_j+} and averaging over the $x$ coordinate, we obtain 
\begin{widetext}
	\begin{equation}
	\delta j_+  =-{e^3\over 2m^2} \Xi(P_{\rm L1}+i{P}_{\rm L2})\int \dd\varepsilon g(\varepsilon)  \varepsilon^2 \left(\tau_c^- \right)'   \tau_{c2}   \left[(\tau_{1\omega}^- + \tau_{1,-\omega}^-) f_0' \right]'.
	\end{equation}
\end{widetext}
Integrating over $\varepsilon$  we get
\begin{equation}
\delta j_+  =-{e^3 \over 2m^2}\Xi(P_{\rm L1}+i{P}_{\rm L2}) \left[g  E_\text{F}^2 \left(\tau_c^- \right)' \tau_{c2} \right]'\qty(\tau_{1\omega}^- + \tau_{1,-\omega}^-).
\end{equation}

According to the same arguments as at calculation of the $EEV$-contribution (see the previous susbsection), the maximal result comes from 
$(\tau_c^-)''$:
\begin{equation}
\label{d_j_fin}
\delta j_+  =-\Xi{e^3 E_\text{F}^2 \over \pi \hbar^2 m} (\tau_c^-)''R(P_{\rm L1}+i{P}_{\rm L2}).
\end{equation}
Here
\begin{equation}
R = \tau_{c2}(\tau_{1\omega}^- + \tau_{1,-\omega}^-).
\end{equation}

\subsubsection{Total ratchet current}
\label{App_total}

A sum of $j_+$ from Eq.~\eqref{j_fin} and $\delta j_+$ from Eq.~\eqref{d_j_fin} yields the total ratchet current in the form
\begin{align}
\label{j_tot_fin}
& j_x+ij_y  =  -\Xi{e^3 E_\text{F}^2 \over \pi \hbar^2 m} (\tau_c^-)''
\biggl[Q_+ + iT_0 \\ &  + (iT_L + R)(P_{\rm L1}+i{P}_{\rm L2})
+ (iT_\text{C} - Q_-)P_\text{C}  \biggr] 
.\nonumber
\end{align}
Substituting 
\begin{equation}
(\tau_c^-)''=\delta_c \left( {2\pi \over \hbar\omega_c} \right)^2    {\tau_1\over(1-i\omega_c\tau_1)^2},
\end{equation}
and finding real and imaginary parts of Eq.~\eqref{j_tot_fin}, we obtain the components of the total ratchet current. They have the form of Eq.~\eqref{j_tot} with the dimensionless coefficients $D_i$ given by
\begin{align}
\label{D_i_DD}
D_0={Q_+ + iT_0 \over\tau_{\rm tr}^2 (1-i\omega_c\tau_{\rm tr})^2},
\quad
D_{L1}={iT_L + R \over\tau_{\rm tr}^2 (1-i\omega_c\tau_{\rm tr})^2}, \nonumber
\\
D_{L2}=iD_{L1}, \quad D_C={iT_\text{C} - Q_- \over\tau_{\rm tr}^2 (1-i\omega_c\tau_{\rm tr})^2}.
\end{align}

\subsubsection{Resonant approximations}

Let us simplify the obtained expressions in vicinity of the CR harmonics assuming $\omega\tau_{1,2}, \omega_c\tau_{1,2} \gg 1$.

First we examine the main CR harmonics $\abs{\omega_c} \approx \omega$. 
For $\omega_c>0$ we have
\begin{multline}
T_0=T_\text{C}={\tau_{1\omega}^+\over 2\omega}, \qquad T_L = -{\tau_{1,-\omega}^-\over 2\omega},
\\ 
Q_\pm = \pm Q_{-\omega} = \pm \tau_{1,-\omega}^- \tau_{2,-\omega}^-,
\qquad R=\tau_{c2}\tau_{1,-\omega}^-.
\end{multline}
Introducing  $\epsilon=(\omega-\omega_c)\tau_1$ we obtain
\begin{equation}
T_0={\tau_1/(2\omega)\over 1-i\epsilon},
\end{equation}
\begin{equation}
T_L = -T_0^*, \qquad R=Q_+=iT_0^*, \qquad Q_-=-iT_0^*.
\end{equation}
Then we get from Eq.~\eqref{j_tot_fin} $j_y \gg j_x$,
\begin{equation}
j_y(\omega \approx  {\omega_c}) = \Xi{e^3 E_\text{F}^2 \over \pi \hbar^2 m} (\tau_c^-)'' {\tau_1\over \omega}
{1+ P_\text{C} \over 1+\epsilon^2}.
\end{equation}
Making the same analysis for $\omega_c<0$ we again obtain $j_y \gg j_x$. For both magnetic field directions $\omega_c><0$ we get ($\epsilon=(\omega-\abs{\omega_c})\tau_1$)
\begin{equation}
\label{DD_resonance}
j_y(\omega \approx \pm {\omega_c}) = \Xi{4\pi e^3 E_\text{F}^2 \over  \hbar^4 m \omega^5} \delta_c
{P_\text{C} \pm 1\over 1+\epsilon^2}.
\end{equation}

Now we turn to the resonance $\omega = \pm 2\omega_c$. In this case we have
\begin{equation}
T_0, T_L, T_\text{C}, R \ll Q_{\pm} 
={i\tau_2\over \omega } \left[ 
\begin{array}{l}
\mp {1\over 1+ i\epsilon_2} \quad \omega_c>0,\\
{1\over 1- i\epsilon_2} \quad \omega_c<0,
\end{array}
\right. 
\end{equation}
where $\epsilon_2= (\omega\mp 2\omega_c)\tau_2$.
This yields
\begin{equation}
\label{DD_2_omega_c_res}
j_+(\omega \approx  \pm {2\omega_c}) = -\Xi{4\pi e^3 E_\text{F}^2 \tau_2\over  \hbar^4 m \omega^5\tau_1} \delta_c
{1\pm P_\text{C} \over 1+\epsilon_2^2} (\epsilon_2 \pm i).
\end{equation}
In this resonance, both components of the ratchet current are present. The $x$- and $y$-components depend on frequency as an antisymmetric and symmetric contours, respectively.

\subsection{General equations for HD regime}
\label{AppHD}

General equation for dc current excited by radiation with arbitrary polarization  within  HD approximation  looks (see details of calculations in Ref.~\cite{Sai2021}):

\be
\label{supp-J}
\begin{aligned}
\bm j	&= \delta_{\rm c}\left(\frac{2\pi E_{\rm F}}{\hbar \omega_{\rm c}}\right)^2 \frac{2\tilde j_0 \gamma^4}
	{\left|\omega_{\rm c}^2 - (\omega-i \gamma)^2\right|^2 (\gamma^2+\omega^2) (\gamma^2+\omega_{\rm c}^2)|D_{\omega q}|^2  }
	\\
	&\times \left(  P_0\hspace{0.2mm} \m A_0+ P_{\rm L1}\hspace{0.2mm} \m A_{\rm L1}
	+ P_{\rm L2}\hspace{0.2mm} \m A_{\rm L2}+ P_{\rm C} \hspace{0.2mm}\m A_{\rm C}   \right).
\end{aligned}
\ee

Here the factor
\be
D_{\omega q}=\omega (\omega+ i\gamma) -q^2s^2-\omega_{\rm c}^2 \frac{\omega}{\omega+ i\gamma}
\ee
in the denomonator is responsible for the MP resonance, while factor  $\left|\omega_{\rm c}^2 - (\omega-i \gamma)^2\right|^2$  describes CR. The dependence of the response on the radiation polarization is
encoded in the vectors $$\m A_i =\left[
\begin{array}{c}
A_i^x \\
A_i^y \\
\end{array}
\right]$$  (here $i=0,{\rm L1, L2,C} $) given by
\begin{align}
\label{a0}
\m A_0 & = 2 \omega^2\omega_{\rm c} \left|\omega_{\rm c}^2 - (\omega-i \gamma)^2\right|^2
\left[
\begin{array}{c}
-\omega_{\rm c} \\
\gamma \\
\end{array}
\right]
\\
&+ s^2q^2 \left[
\begin{array}{c}
(\gamma^2+2\omega_{\rm c}^2)(\gamma^2+\omega^2)^2 +(\gamma^2-2\omega^2)\omega_{\rm c}^4 \\
\gamma \omega_{\rm c} \left[ \gamma^4-\omega^4 +\omega_{\rm c}^4 + 2 \omega_{\rm c}^2(\gamma^2+2\omega^2)  \right]\\
\end{array}
\right],
\nonumber
\\
\label{aL1}
\m A_{\rm L2} & =s^2q^2 \gamma( \gamma^2+\omega^2+\omega_c^2)\left[
\begin{array}{c}
-\omega_{\rm c} (3 \gamma^2 +\omega^2-\omega_{\rm c}^2)\\
\gamma ( \gamma^2 +\omega^2- 3\omega_{\rm c}^2) \\
\end{array}
\right],
\\
\label{aL2}
\m A_{\rm L1}& = \m A_{\rm L2} \times \hat{\bm z}
,
\\
\label{aC}
\m A_{\rm C} & =\omega (\omega^2+\gamma^2+\omega_{\rm c}^2)
\left|\omega_{\rm c}^2 - (\omega-i \gamma)^2\right|^2
\left[
\begin{array}{c}
\omega_{\rm c} \\
-\gamma \\
\end{array}
\right]
+ \frac{\omega}{\gamma}\m A_{\rm L2}.
\end{align}
We notice that at small $q$ vectors  $\m A_{\rm L1}$ and $\m A_{L2}$ are small, $\propto q^2.$
In other words, dependence on the direction of the linear polarization appears only due to the plasmonic effects. Dimensionless coefficients $\tilde D_i$ entering Eq.~\eqref{j_tot_HD} are expressed in terms of $\mathbf A_i$ as follows:
\be
\tilde D_i=\frac{2\gamma^4 (A_i^x+ i A_i^y)}{\left|\omega_{\rm c}^2 - (\omega-i \gamma)^2\right|^2 (\gamma^2+\omega^2) (\gamma^2+\omega_{\rm c}^2)|D_{\omega q}|^2  }.
\label{tildeD }
\ee

\bibliography{all_lib.bib}

\end{document}